\documentclass[aps, prl, twocolumn, 10pt, superscriptaddress, floatfix]{revtex4-2}
\usepackage{amsmath,amssymb}
\usepackage{braket}
\usepackage{mathtools}
\usepackage{graphicx,diagbox}
\usepackage{xcolor}
\usepackage{ulem}
\usepackage{xspace}
\usepackage{upquote}
\usepackage{csquotes}
\usepackage{tabularx,makecell}
\usepackage{bm,dsfont}

\DeclareUnicodeCharacter{0650}{\unskip}

\usepackage[bookmarks=true,colorlinks=true,urlcolor=blue,linkcolor=blue,citecolor=blue,breaklinks]{hyperref}
\usepackage{booktabs}

\graphicspath{{figures/}}

\begin{document}

\title{ِAnalytical Model for Atomic Relaxation in Twisted Moir\'e Materials}

\author{Mohammed M. Al Ezzi}
\affiliation{Department of Materials Science and Engineering, 
National University of Singapore, 9 Engineering Drive 1, 
Singapore 117575}
\affiliation{Centre for Advanced 2D Materials, National University of Singapore, 6 Science Drive 2, Singapore 117546}
\affiliation{Department of Physics, Faculty of Science, National University of Singapore, 2 Science Drive 3, Singapore 117542}
\author{Gayani N. Pallewela}
\affiliation{Centre for Advanced 2D Materials, National University of Singapore, 6 Science Drive 2, Singapore 117546}
\author{Christophe De Beule}
\affiliation{Department of Physics and Astronomy, University of Pennsylvania, Philadelphia, Pennsylvania 19104, USA}
\author{E. J. Mele}
\affiliation{Department of Physics and Astronomy, University of Pennsylvania, Philadelphia, Pennsylvania 19104, USA}
\author{Shaffique Adam}
\affiliation{Department of Materials Science and Engineering, 
National University of Singapore, 9 Engineering Drive 1, 
Singapore 117575}
\affiliation{Department of Physics and Astronomy, University of Pennsylvania, Philadelphia, Pennsylvania 19104, USA}
\affiliation{Department of Physics, Washington University in St. Louis, St. Louis, Missouri 63130, United States}

\date{\today} 

\begin{abstract}
\normalsize
By virtue of being atomically thin, the electronic properties of heterostructures built from two-dimensional materials are strongly influenced by atomic relaxation. The atomic layers behave as flexible membranes rather than rigid crystals. Here we develop an analytical theory of lattice relaxation in twisted moir\'e materials. We obtain analytical results for the lattice displacements and corresponding pseudo gauge fields, as a function of twist angle. We benchmark our results for twisted bilayer graphene and twisted WSe$_2$ bilayers using large-scale molecular dynamics simulations. Our \textit{single-parameter} theory is valid in graphene bilayers for twist angles $\theta ~\gtrsim 0.7^\circ$, and in twisted WSe$_2$ for $\theta ~\gtrsim 1.6^\circ$. We also investigate how relaxation alters the electronic structure in twisted bilayer graphene, providing a simple extension to the continuum model to account for lattice relaxation.
\end{abstract}

\maketitle

\emph{\it Introduction} --- Since the first paper by dos Santos, Peres and Castro Neto \cite{dos2007graphene}, the community has been fascinated by the modification of the electronic properties of two-dimensional (2D) materials using twist angle \cite{trambly2010localization,mele2010commensuration}. Twisted 2D materials have observable moiré patterns that depend on twist angle \cite{li2010observation,wong2015local}, lattice symmetry \cite{liu2018tailoring,plumadore2020moire}, and lattice mismatch \cite{yankowitz2012emergence}. These moiré patterns in turn strongly modify the electronic properties in ways that can be observed in STM \cite{luican2011single,wong2015local}, transport \cite{yankowitz2012emergence,ponomarenko2013cloning,dean2013hofstadter,hunt2013massive,yankowitz2018dynamic}, and ARPES \cite{sprinkle2009first,ohta2012evidence,sato2021observation,jiang2023revealing}.  Following the seminal experimental observation flat bands in 2018 \cite{cao2018unconventional}, the field experienced a surge of exploration and discovery.  There is now a large family of new artificial superlattices with tunable superlattice periods $\sim10$ nm hosting a rich assortment of interacting electronic phenomena \cite{cao2020tunable,liu2020tunable, xu2021tunable,chen2021electrically,polshyn2020electrical, park2021tunable, hao2021electric,al2024topological,park2022robust, lee2021tunable, mullan2023mixing, waters2023mixed, tang2020simulation, wang2020correlated, regan2020mott, cai2023signatures, zhao2023time}.  

One wrinkle in the moiré story is that 2D atomic layers are not rigid, but flexible electronic membranes.  This was known in early experimental studies \cite{woods2014commensurate,yamamoto2012princess}. Theoretical treatments considering the electronic and mechanical properties on equal footing soon followed \cite{jung2015origin,san2014electronic,san2014spontaneous}. At small twist angles, small atomic displacements from the rigid twisted configuration give large gains in the electronic potential energy
at a small cost in elastic energy.  Since lattice relaxation is expected to strongly modify the electronic structure, the original claims of flat bands \cite{bistritzer2011moire} were met with some skepticism. It was only after the experimental observations that relaxation effects were considered in magic angle twisted bilayer graphene (tBG). As expected, relaxation effects strongly modify the electronic structure. Remarkably, relaxation actually further flattens and isolates the lowest energy moir\'e bands \cite{nam2017lattice,carr2019exact,guinea2019continuum} confirming that atomic relaxation is an important ingredient to understand the observed superconductivity and correlated insulator states.  

In this Letter we propose a symmetry-based and fully analytical physically motivated theory for atomic relaxation. By benchmarking our results with \textsc{lammps} molecular dynamics simulations (involving numerical calculations with more than 4 million atoms), we show that our theory is valid for a wide range of twist angles for both tBG, including at the magic angle, and parallel-stacked twisted transition metal dichalcogenides (tTMDs) homobilayers, i.e., where corresponding atoms in two layers are aligned. This enables us to propose an effective electronic model for twisted bilayer graphene that fully captures the effects of lattice relaxation on observables including band width, Fermi velocity and pseudomagnetic fields. We note that the success of the original rigid model \cite{dos2007graphene, bistritzer2011moire} was largely due to its conceptual clarity and computational simplicity making it broadly accessible. Our simple analytic extension to this model to account for atomic relaxation retains all of these advantages. 
\begin{figure}
    \centering    \includegraphics[width=.95\linewidth]{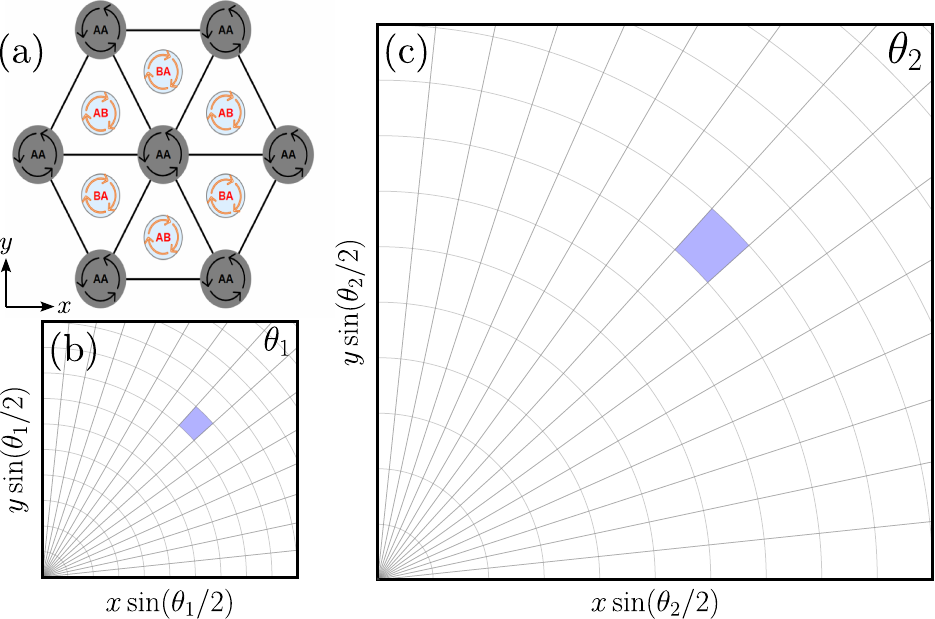}
    \caption{Illustration of the analytical model for lattice relaxation in twisted moir\'es. (a) Lattice relaxation prompts neighborhoods of energetically favorable stacking domains to counter-rotate against external twisting, to locally restore the favorable stacking. Simultaneously, atoms in unfavorable domains co-rotate with the twist. This suggests that $\nabla \cdot \bm u = 0$ where $\bm u(\bm r)$ is the in-plane displacement field. (b, c) Conformal map between different moir\'e cells. While coordinates scale as $1/\sin(\theta/2)$, the displacement field, proportional to the van der Waals force in a small area, scales as $1/\sin^2(\theta/2)$.}
    \label{fig:cartoon}
\end{figure}

The importance of lattice reconstruction is determined by the dimensionless quantity $V_1 / [ \mu \sin^2(\theta/2) ]$, where $V_1$ is the van der Waals energy scale and $\mu$ is a Lam\'e elastic coefficient \cite{nam2017lattice}.  Using accepted values \cite{PhysRevB.98.195432, enaldiev2023dislocations}, we anticipate that relaxation will be an order-of-magnitude stronger in tTMDs compared to tBG; since lattice relaxation effects are expected to be important for $\theta \lesssim 1^\circ$ in tBG, we expect it to be important already at larger angles for tTMDs. Previous studies of lattice relaxation have relied on numerical solutions of elasticity theory \cite{jung2015origin, nam2017lattice}, effective models fitted to DFT \cite{carr2019exact}, or large-scale classical atomic molecular dynamics calculations \cite{guinea2019continuum}. Here we provide a largely analytical approach which is fully consistent with prior numerical works over a wide range of the parameter space and provides both additional insights and a flexible method for treating relaxation.

\begin{figure*}[!t]
    \centering
    \includegraphics[width=.93\linewidth]{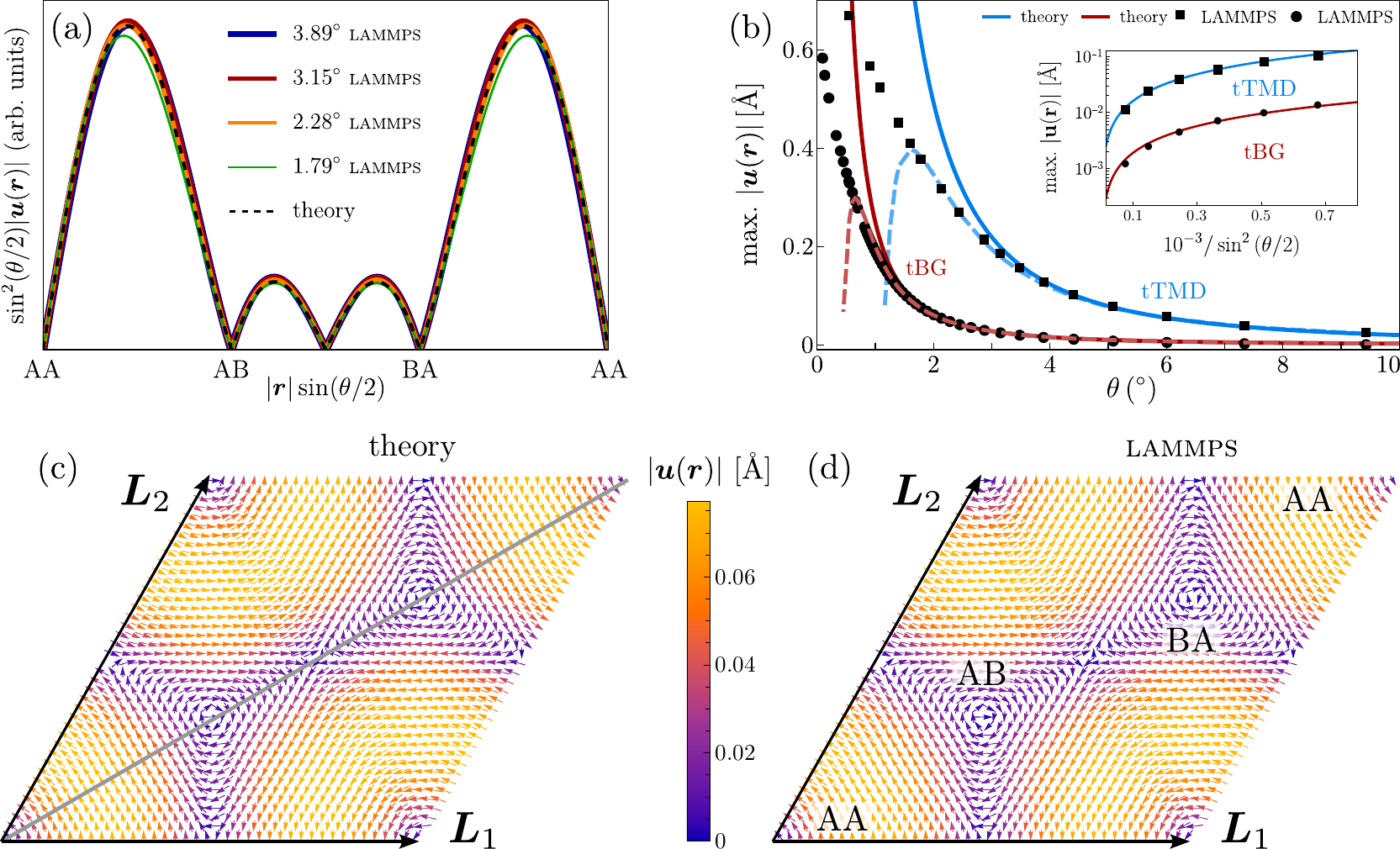}
    \caption{Validation of theory. (a) Magnitude of the in-plane displacement field $\bm u(\bm r)$ due to lattice relaxation along the diagonal line in panel (c) for different twist angles in tBG. The coordinate is scaled by the moir\'e period while the displacement is scaled by the moir\'e cell area. The \textsc{lammps} data (solid lines) collapses to the same curve for sufficiently large twist angles after scaling as predicted from theory (dashed line). (b) Maximum $|\bm u(\bm r)|$ as a function of twist angle, comparing \textsc{lammps} (data points) to the theory for twisted bilayer graphene and twisted bilayer WSe$_2$ near parallel stacking.  The theory corresponds to Eq.\ \eqref{eq:main} with $\alpha_1 = c_1 (L/a)^2$ (solid) and $\alpha_1 = c_1 (L/a)^2 - c_1^2 (L/a)^4$ (dashed). The single adjustable parameter is $c_1 = 6.5 \times 10^{-5}$ for tBG and $c_1 = 3.9 \times 10^{-4}$ for tWSe$_2$ (c) In-plane displacement field of the top tBG layer calculated with Eq.\ \eqref{eq:main} for a twist angle $\theta = 1.79^\circ$. (d) In-plane displacement field computed with \textsc{lammps} for the same twist angle as (c).}
    \label{fig2}
\end{figure*}


\emph{\it{Analytical Model}} --- We begin by considering atomic relaxation in twisted bilayer graphene and extend our results to other twisted homobilayers later. Without relaxation, the geometry of tBG is obtained by starting from a fixed stacking (e.g., AA stacking) and introducing a relative twist angle $\theta$ between layers. This results in the rigid configuration of the twisted system \cite{dos2012continuum}. However, the 2D atomic sheets act more like elastic membranes, gaining energy by allowing atoms to relax. The relaxed atomic positions of the top and bottom layers are $\bm r \pm \left[ \bm u(\bm r) + h(\bm r) \hat z \right]$, respectively, where $\bm r$ are the rigid in-plane coordinates, and $\bm u(\bm r)$ and $h(\bm r)$ are the relative in-plane and out-of-plane displacement fields due to lattice relaxation, respectively.  
As discussed in the Supplemental Material (SM) \footnote{See Supplementary Material at [insert link] for more details}  (which
includes Refs. \cite{zou2018band,mchugh2023_moire,JeilFF,garcia2021_full,dresselhaus_group_2007,balents_general_2019,amorim_novel_2016,landau_theory_1986,kang2023b,moon2013_optical,THOMPSON2022108171,Donald:WBrenner_2002,doi:10.1021/acs.jpcc.8b10392, Jiang_2015}), for the angles we consider, in-plane homostrain and buckling are negligible.

We first approach the relaxation problem with a hypothesis based on physical intuition, and then confirm it with a symmetry analysis. The relative rotation of the layers induces variation in the local stacking registry, forming a triangular pattern of high-symmetry stacking points, as shown in Fig.\ \ref{fig:cartoon}(a). The AB and BA regions are degenerate energy minima, whereas the AA regions are energy maxima \cite{alden2013strain}. The AB and BA regions maintain their low-energy configuration by resisting rigid rotation and tend to locally counter-rotate against the global rigid rotation. Conversely, to decrease their energy, the AA regions tend to rotate further in the same direction, resulting in an additional local rotation, as illustrated in Fig.\ \ref{fig:cartoon}(a). This gives rise to a triangular lattice of vortices in $\bm u$, with opposite vorticity at the AB and BA regions compared to the AA regions. This physical picture leads us to hypothesize that lattice relaxation will be primarily rotational, resulting in an incompressible (solenoidal) displacement field:
\begin{equation} \label{eq:incompr}
    \nabla \cdot \bm u(\bm r) = 0,
\end{equation}
\begin{table}
    \begin{tabular}{c | c | c | c | c}
        \Xhline{1pt}
        $m$ & $g/g_1$ & $\alpha_m$ & $\beta_m$ & $h_m$ \\
        \hline
        $1$ & $1$ & $\mathds R$ & $0$ & $\mathds R$ \\ 
        $2$ & $\sqrt{3}$ & $\mathds R$ & $0$ & $\mathds R$ \\
        $3$ & $2$ & $\mathds R$ & $0$ & $\mathds R$ \\
        $4,5$ & $\sqrt{7}$ & $\mathds R$ & $\mathds R$ & $\mathds R$ \\
        \Xhline{1pt}
    \end{tabular}
    \caption{Allowed values of the in-plane ($\alpha_m$ and $\beta_m$) and out-of-plane ($h_m$) Fourier components of the displacement fields, for $D_6$ symmetry for the first five moir\'e stars indexed by $m$ with $g_1 = 4\pi / \sqrt{3} L$.}
    \label{tab:symmetry}
\end{table}

We now show that symmetries require that Eq.\ \eqref{eq:incompr} is an exact constraint for sufficiently smooth relaxation patterns. Since the rigid configuration is adiabatically connected to the relaxed one, the displacement field $\bm u(\bm r)$ should obey the same symmetries of the moir\'e and vary slowly on the atomic scale. Hence it can be expanded as
\begin{equation}
    \bm u(\bm r) = \sum_{\bm g} \bm u_{\bm g} e^{i\bm g \cdot \bm r}, 
\end{equation}
where the sum runs over moir\'e reciprocal vectors and $\bm u_{\bm g}$ are in-plane Fourier components.  A Helmholtz decomposition in terms of transverse and longitudinal parts gives
\begin{equation}
    \bm u_{\bm g} = \frac{a}{L} \frac{\alpha_{\bm g} \hat z \times \bm g + \beta_{\bm g} \bm g}{ig^2},
\end{equation}
where the $\alpha_{\bm g}$ ($\beta_{\bm g}$) are dimensionless and correspond to rotational (volumetric) displacements, $a$ is the lattice constant of graphene, and $L$ is the moir\'e lattice constant. While these coefficients are \textit{a priori} unknown, they are constrained by symmetry. For example, an in-plane symmetry $\mathcal S$ requires that $\mathcal S \bm u(\bm r) = \bm u(\mathcal S \bm r)$ giving 
$\alpha_{\mathcal S\bm g} = \det(\mathcal S) \alpha_{\bm g}$ and $\beta_{\mathcal S\bm g} = \beta_{\bm g}$. Taking into account the emergent $D_6$ symmetry of the moir\'e, we find constraints on the $\alpha_m$ and $\beta_m$ where $m$ labels moir\'e reciprocal stars. Here each star contains six reciprocal vectors closed under $\mathcal C_{6z}$ rotations \cite{Note1}. The allowed values for the first five stars are shown in Table \ref{tab:symmetry}. Importantly, we find that the $D_6$ symmetry strongly suppresses volumetric contributions, which in fact vanish up to the third star. Restricting to the first star, we obtain
\begin{equation} \label{eq:main}
    \bm u(\bm r) = \alpha_1 \frac{\sqrt{3} a}{2\pi} \sum_{i=1}^3 \hat z \times \hat g_i \sin(\bm g_i \cdot \bm r),
\end{equation} 
where the sum runs over three reciprocal vectors of the first star related by $\mathcal C_{3z}$ and $\alpha_1 = \alpha_{\bm g_1} = \alpha_{\bm g_2} = \alpha_{\bm g_3}$. Equation \eqref{eq:main} is one of the main results of our work. It gives an analytical expression for $\bm u(\bm r)$ with a single dimensionless parameter $\alpha_1$ that can be determined either from theory and simulations (as we do below), or experiment.

Next we investigate $\alpha_1(\theta)$. In Fig.\ \ref{fig:cartoon} we observe that the geometry of moir\'e cells corresponding to different twist angles can be mapped: a point $\bm r_1$ in a moir\'e cell ($\theta_1$) is mapped to $\bm r_2$ in another moir\'e cell ($\theta_2$) with identical local environments, according to $\bm r_1 \sin(\theta_1/2) = \bm r_2 \sin(\theta_2/2)$. Thus, coordinates scale as $1/\sin(\theta/2)$. Likewise, we can map areas between different moir\'es. If we assume that the van der Waals interaction varies slowly on the atomic scale, the net force in an area patch scales as $1/\sin^2(\theta/2)$. For small displacements, the displacement field, written in fractional coordinates $\bm r^\prime$,  is proportional to the force giving
\begin{equation}
    \sin^2 \left( \tfrac{\theta_1}{2} \right) \bm u_1 \left( \bm r^\prime / \sin \tfrac{\theta_1}{2} \right) = \sin^2 \left( \tfrac{\theta_2}{2} \right) \bm u_2 \left( \bm r^\prime / \sin \tfrac{\theta_2}{2} \right), \label{eq:scaling}
\end{equation}
which implies that $\alpha_1(\theta) = c_1 / [4 \sin^2(\theta/2)]$ where we defined a materials constant $c_1$ that quantifies the strength of lattice relaxation.
Using elastic theory, with the \textit{ansatz} from Eq.\ \eqref{eq:main}, we find $c_1 = V_1 / \mu$ \cite{Note1}.  

An immediate consequence is that the displacement field divided by the moir\'e cell area as a function of $\bm r / L$ is independent of twist angle. We test this collapse using \textsc{lammps} molecular dynamics simulations, as shown in Fig.\ \ref{fig2}(a). 
Beyond the collapse of the numerical data for different twist angles, we find excellent agreement with the model [dashed line in Fig.\ \ref{fig2}(a)]. Next, we use the \textsc{lammps} data to fix $c_1$, the single material-dependent parameter of the theory. By fitting the data for large twist angles, as shown in Fig.\ \ref{fig2}(b), we find $c_{1,\text{tBG}} \approx 6.5 \times 10^{-5}$ for tBG and $c_{1,\text{tWSe$_2$}} \approx 3.9 \times 10^{-4}$ for parallel-stacked bilayer WSe$_2$ (since TMDs lack $\mathcal C_{2z}$ symmetry, one distinguishes between rotating away from parallel and antiparallel stacking). As anticipated, we find $c_{1,\text{tWSe$_2$}} \gg c_{1,\text{tBG}}$ since TMDs are both elastically softer and have
a larger energy difference between AA (metal on metal) and AB (metal on chalcogen) stacking \cite{carr2018relaxation}. Interestingly, even though moir\'es of twisted homobilayer TMDs only have $D_3$ symmetry, the stacking-fault energy for parallel stacking has $D_6$ symmetry as a function of atomic disregistry. Such an emergent symmetry is a generic feature of moir\'e materials \cite{angeli2021_gamma} and explains why our theory works well for both tBG and parallel-stacked homobilayer tTMDs.
\begin{figure}
    \centering
    \includegraphics[width=\linewidth]{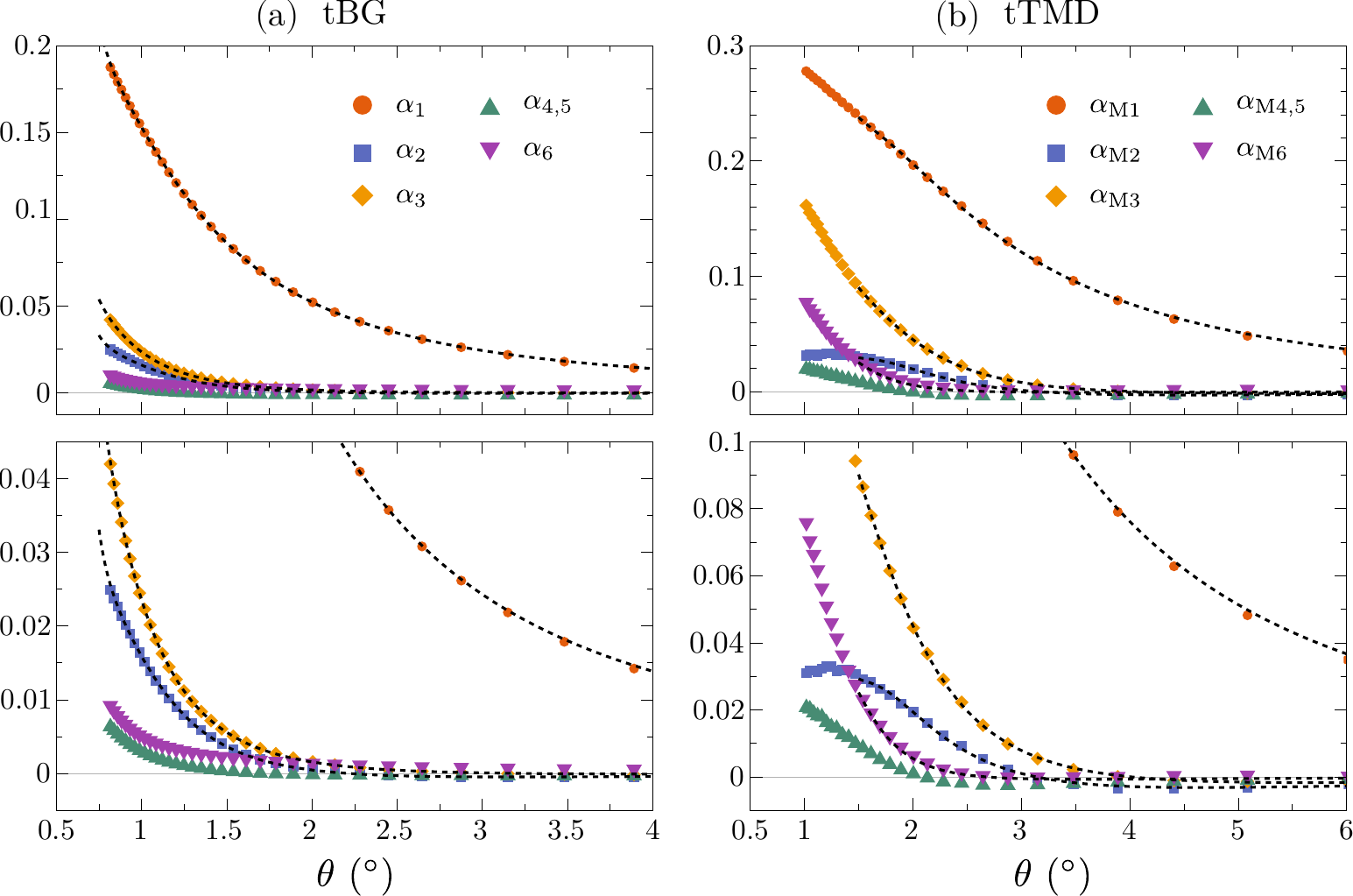}
    \caption{Rotational components $\alpha_m$ of the in-plane displacement field $\bm u(\bm r)$ for different stars $m$ as a function of twist angle for (a) tBG and (b) twisted parallel-stacked WSe$_2$ where $\bm u(\bm r)$ is obtained from the W atoms. Dashed lines give the series of Eq.\ \eqref{eq:alphaseries} up to fourth order (fitting parameters given in SM).}
    \label{fig:mainalpha}
\end{figure}

Having fixed the parameter $c_1$ with \textsc{lammps}, we now compare the full profile $\bm u(\bm r)$ between theory and simulations for tBG. This is shown for $\theta = 1.79^\circ$ in Fig.~\ref{fig2}(c) and (d). We see that the displacement fields are nearly identical. Remarkably, for tBG the first-order theory is valid for $\theta \gtrsim 1^\circ$ and works even at the magic angle. For tTMDs, on the other hand, lattice relaxation is stronger, giving larger displacements for the same twist angle. For example, the displacements for tWSe$_2$ near $4^\circ$ have similar magnitudes as those for tBG near $1^\circ$, see Fig.\ \ref{fig2}(b). As the twist angle decreases below these values, the first-order theory starts to deviate from \textsc{lammps}. We attribute this breakdown in simple scaling to nonlinear contributions to the van der Waals energy. In general, one has the expansion
\begin{equation} \label{eq:alphaseries}
    \alpha_i = \sum_{n=1}^\infty \frac{A_{in}}{\left( 4 \sin^2 \tfrac{\theta}{2} \right)^n},
\end{equation}
where $i$ labels the reciprocal star and the $A_{in}$ are determined either from continuum elasticity, \textsc{lammps} simulations, or DFT. In the first case, we minimize the total potential energy $U[\bm u(\bm r)] = U_\text{elastic} + U_\text{vdW}$ subject to $D_6$ symmetry. Here we assume that out-of-plane displacements are small compared to in-plane ones. Expanding the stacking-fault energy in lowest order of $|\bm u|/a$,
\begin{equation}
    U \sim \frac{a^2}{L^2} \sum_{\bm g} \left[    \mu | \alpha_{\bm g} |^2 + \left( \lambda + 2 \mu \right) | \beta_{\bm g} |^2 \right] - 2 \sum_{\bm g} V_{\bm g} \alpha_{-\bm g},
\end{equation}
where $\mu$ and $\lambda$ are Lam\'e parameters and $V_{\bm g}$ are Fourier components of the stacking-fault energy. The energy is minimized by $\alpha_{\bm g} = L^2 V_{\bm g} / \mu a^2$ and $\beta_{\bm g} = 0$. Hence, the in-plane displacement field to lowest order is purely rotational in \textit{twisted} moir\'e materials regardless of commensuration or symmetry constraints, consistent with experimental observations \cite{kazmierczak2021_strain}. As discussed earlier, symmetry can further suppress volumetric terms which is shown explicitly for $D_6$ in Table \ref{tab:symmetry}. Perturbatively, we have $A_{in} = \delta_{n1} V_i / \mu$ with leading-order corrections $A_{22} = A_{32} = -A_{12} = c_1^2$ (see SM).
Furthermore, keeping terms up to fourth order in Eq.\ \eqref{eq:alphaseries}, we find excellent agreement with \textsc{lammps} for all data shown for tBG and down to $1.4^\circ$ for tWSe$_2$. This is shown in Fig.\ \ref{fig:mainalpha} for the first three stars. Near perfect agreement between \textsc{lammps} and DFT can be obtained using the GAP20 potential \cite{leconte2022relaxation} for which we find $c_{1,\text{tBG}} = 4.5 \times 10^{-5}$. However, this force field becomes computationally challenging when $\theta \lesssim 1^\circ$.

We briefly mention that the out-of-plane hetero displacements can be understood from a local-stacking picture where the interlayer distance follows the stacking type defined by the in-plane atomic positions. In the first-star approximation we find $h_0 = \left( h_\text{AA} + 2 h_\text{AB} \right) / 6 - \left( h_\text{AA} - h_\text{AB} \right) c_1 / 6 \sin(\theta/2)^2$ and $h_1 = \left( h_\text{AA} - h_\text{AB} \right) \left[ 1 - c_1 / 2 \sin(\theta/2)^2 \right] / 6$ \cite{Note1}.
 
\emph{\it{Effective Relaxed Electronic Model}} --- Armed with the analytical theory for $\bm u(\bm r)$, we can now incorporate relaxation into a continuum model for the low-energy bands of tBG \cite{bistritzer2011moire}. 
In the continuum model, the low-energy Dirac fermions of the two graphene layers are coupled by an interlayer moir\'e potential that is expanded in successive harmonics \cite{Note1}. 
Close to the magic angle, the moir\'e potential is dominated by the first harmonic. We propose a continuum model valid for $\theta \gtrsim 1^\circ$ with a modified interlayer moir\'e coupling,
\begin{equation}
    T(\bm r) = \sum_{j=1}^3 T_j e^{i \bm q_j \cdot \bm r}, \label{4thgen}
\end{equation}
where $T_j = e^{-i(j-1)\pi \sigma_z/3} \left( w_1 e^{i\phi \sigma_z} + w_2 \sigma_x \right) e^{i(j-1)\pi \sigma_z/3}$ with $\sigma_{x,y,z}$ Pauli matrices, and where $\bm q_i$ are moir\'e tunnelling vectors. Here $w_1$ and $w_2$ are tunnelling amplitudes between equal and opposite sublattices, respectively.
We introduce a new symmetry-allowed parameter $\phi$ 
(also considered recently in Ref.\ \cite{kang2023}) that encodes information about the lattice relaxation \cite{Note1}. In the absence of relaxation, within a two-center approximation, $w_1 = w_2$ and $\phi = 0$ and only one moir\'e star is necessary \cite{dos2007graphene,dos2012continuum,bistritzer2011moire}. In this case, there is no energy gap between the flat bands and remote dispersive bands. Including relaxation, not only is $w_1 \neq w_2$, and the gap becomes finite \cite{nam2017lattice,koshino2018_maximally}, but higher-order shells as well as nonlocal moir\'e terms become important. For this paper, we keep only one star, encoding the information of higher shells in $\phi$. The resulting electronic spectrum \cite{Note1}
matches large-scale relaxed tight-binding calculations \cite{angeli2018emergent, leconte2022relaxation}.

\begin{figure}[!t]
    \centering
    \includegraphics[width=.95\linewidth]{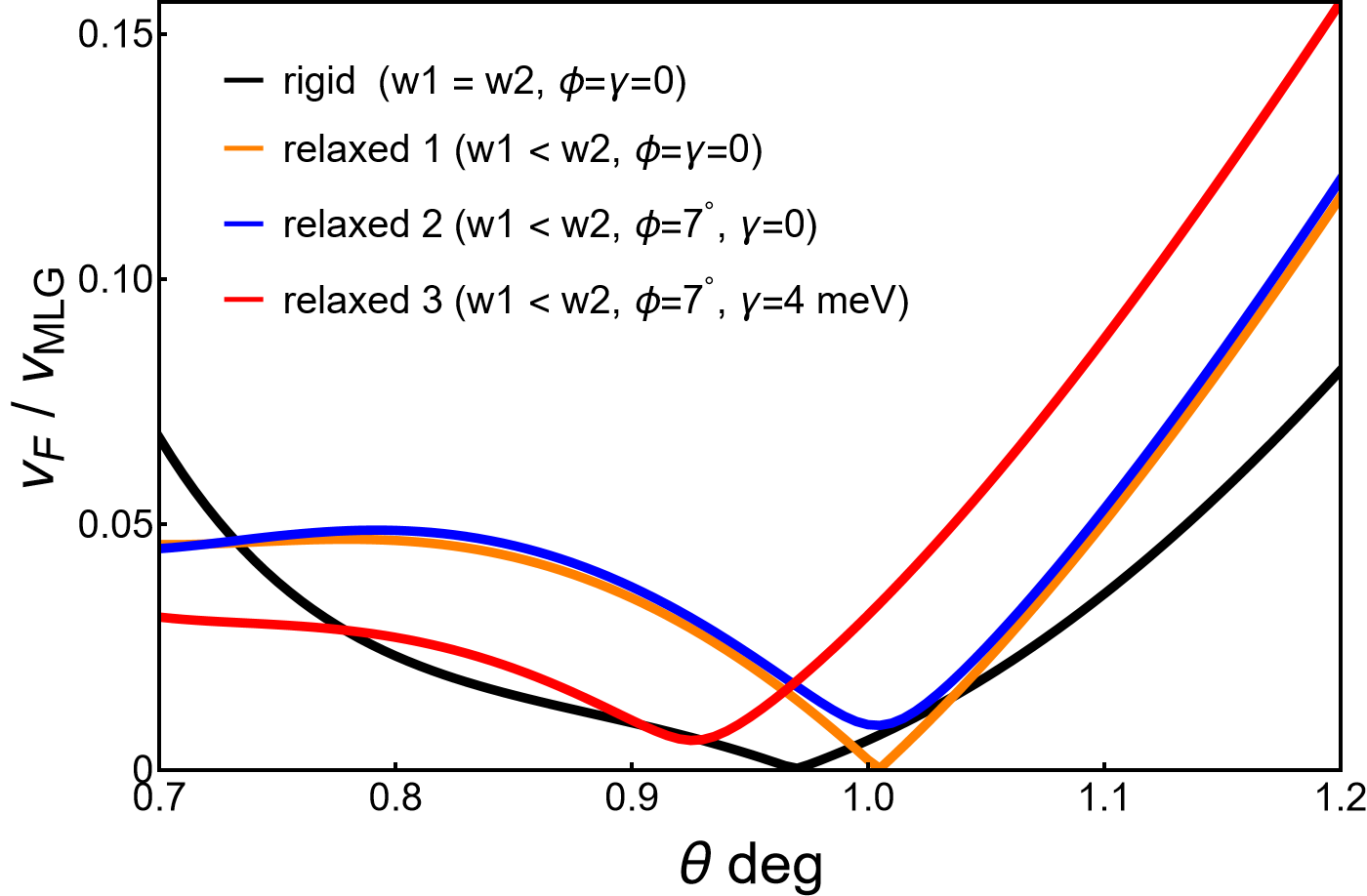}
    \caption{Modification of Fermi velocity $v_F$ due to lattice relaxation in tBG. The black curve corresponds to the rigid case with $w_1 = w_2$ and $\phi = 0$. Taking into account relaxation, we consider $w_1 \neq w_2$ with $\phi=0$ (orange) and finite $\phi$ (blue) where $\phi$ is the AA tunneling phase. The latter makes $v_F$ complex yielding a finite minimum value at the nominal magic angle. Finally, the red curve also includes the pseudomagnetic field. Here $v_F$ is given in units of the monolayer value.}
    \label{fig4}
\end{figure}

Perturbatively, one can show that $\theta_\text{magic} \propto w_2 a / \hbar v_F$ where the magic angle is defined by vanishing Fermi velocity \cite{bistritzer2011moire,Note1}. Since $w_2$ increases with relaxation \cite{carr2019exact}, the magic angle increases. However, relaxation induces strain giving rise to a pseudomagnetic field \cite{vozmediano_gauge_2010} which we find is on the order of $10 \, \text{T}$ near $1^\circ$. The displacement field in Eq.\ \eqref{eq:main} yields a pseudo vector potential $\pm \bm A(\bm r)$ for top and bottom layer, respectively. Here 
$\bm A(\bm r) = [ 2 \gamma(\theta) / ev_F ] \sum_{i=1}^3 \hat z \times \hat g_i \cos(\bm g_i \cdot \bm r)$ where 
$\gamma = \sqrt{3} \hbar v_F L / 4 \pi \ell_0^2$ with $\ell_0 \sim a / \sqrt{2\pi c_1}$ the effective magnetic length \cite{Note1}. Importantly, the pseudomagnetic field reduces the magic angle such that overall the magic angle does not change much. By contrast, the phase $\phi$ makes the Fermi velocity complex and it attains only a finite minimum value as shown in Fig.\ \ref{fig4}.  Close to magic angle $w_2 = 97$ meV, $w_1 = 79$ meV, and $\gamma = 4$ meV \cite{Note1}. 

\emph{\it Conclusion} --- The field of moir\'e materials is rapidly expanding, and there is a need for transparent and physically motivated  models to understand and interpret the large influx of experimental data. In this work, we have proposed an analytical theory for atomic relaxation in twisted moir\'es based on symmetry and scaling arguments. We have benchmarked our model against large-scale molecular dynamics showing remarkable agreement. The model has several important consequences including explicit twist-angle dependence for moir\'e tunneling amplitudes and pseudomagnetic fields, and the shift of the magic angle.
 Our formalism can be extended to other twisted materials with different stacking-fault symmetries, as well as other moir\'es beyond twisting \cite{CDB2024}. 

\emph{\it Note added} --- During the preparation of this manuscript, Ceferino and Guinea submitted a preprint \cite{ceferino2023pseudomagnetic} with a similar model for relaxed twisted bilayer and trilayer graphene.  

\let\oldaddcontentsline\addcontentsline 
\renewcommand{\addcontentsline}[3]{} 
\begin{acknowledgments}
MMAE, GNP, and SA acknowledge financial support from the Singapore National Research Foundation Investigator Award (NRF-NRFI06-2020-0003) and the Singapore Ministry of Education AcRF Tier 2 grant (MOE-T2EP50220-0016). CDB and EJM are supported by the U.S.\ Department of Energy under Grant No.\ DE-FG02-84ER45118. GNP acknowledges National University of Singapore HPC (NUSREC-HPC-00001) and thanks Miguel Dias Costa for assistance with the numerical simulations.
\end{acknowledgments}

\bibliographystyle{unsrt}
\bibliography{references}
\let\addcontentsline\oldaddcontentsline 


\clearpage
\onecolumngrid
\begin{center}
\textbf{\Large Supplemental Material}
\end{center}
 
\setcounter{equation}{0}
\setcounter{figure}{0}
\setcounter{table}{0}
\setcounter{page}{1}
\setcounter{secnumdepth}{2}
\makeatletter
\renewcommand{\thepage}{S\arabic{page}}
\renewcommand{\thesection}{S\arabic{section}}
\renewcommand{\theequation}{S\arabic{equation}}
\renewcommand{\thefigure}{S\arabic{figure}}
\renewcommand{\thetable}{S\arabic{table}}

\tableofcontents
\vspace{1cm}

\twocolumngrid

\section{Displacement fields from symmetry} \label{app:symmetry}

\begin{figure}
    \centering
    \includegraphics[width=\linewidth]{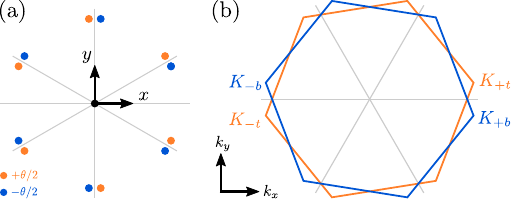}
    \caption{(a) Twisted bilayer graphene with $D_6$ symmetry showing only carbon atoms closest to the twist center (black dot). Symmetry axes are shown in gray. (b) Corresponding momentum space of the two rotated layers.}
    \label{fig:tbgD6}
\end{figure}

In this section, we constrain the displacement fields due to relaxation in twisted bilayer graphene (tBG) using the symmetry of the moir\'e lattice. In particular, we consider commensurate approximants with the twist center at a graphene hexagon center that have the periodicity of the moir\'e lattice. These structures have point group $D_6 = \left< \mathcal C_{6z}, \mathcal C_{2x} \right>$ where $\mathcal C_{6z}$ is a rotation by $\pi/3$ about the $z$ axis and $\mathcal C_{2x}$ is a $\pi$ rotation about the $x$ axis \cite{zou2018band}, as illustrated in Fig.\ \ref{fig:tbgD6}(a). The corresponding commensurate twist angles are given by \cite{dos2012continuum}
\begin{equation}
    \cos \theta_{mr} = \frac{3m^2 + 3mr + r^2/2}{3m^2 + 3mr + r^2},
\end{equation}
with $r=1$.
The displacements fields are defined through the atomic positions
\begin{align}
    \bm r_t & = \bm r_+ + \bm u_t(\bm r_+) + h_t(\bm r_+) \hat z, \\
    \bm r_b & = \bm r_- + \bm u_b(\bm r_-) + h_b(\bm r_-) \hat z,
\end{align}
with $\bm r_\pm = R_{\pm\theta/2} \bm r$ where $\bm r = n_1 \bm a_1 + n_2 \bm a_2 + \bm \delta$ are the atomic positions of monolayer graphene with $n_1,n_2 \in \mathds Z$ and $\bm \delta$ the sublattice position in the graphene cell. Here the origin is placed at a hexagon center and $R_{\pm\theta/2}$ is the rotation matrix for a counterclockwise rotation about the $z$ axis by an angle $\pm\theta/2$. In the absence of lattice relaxation, i.e., the rigid twisted configuration, the in-plane displacement fields $\bm u_{t/b}$ vanish and $h_{t/b}(\bm r) = \pm h_0$ is constant. We further define the displacement fields
\begin{align}
    \bm u_{t/b}(\bm r) & = \overline{\bm u}(\bm r) \pm \bm u(\bm r), \\
    h_{t/b}(\bm r) & = \overline h(\bm r) \pm h(\bm r),
\end{align}
since both the homo ($\overline{\bm u}$ and $\overline h$) and hetero ($\bm u$ and $h$) displacements are symmetry-allowed and transform properly under $D_6$. Physically, $\overline h$ corresponds to buckling of the graphene sheets while $h$ is a breathing mode. To keep the discussion succinct, we focus on the relative displacements $\bm u(\bm r)$ and $h(\bm r)$ and only give the final results for $\overline{\bm u}(\bm r)$ and $\overline h(\bm r)$. Assuming the moir\'e periodicity is preserved after lattice relaxation, we can write 
\begin{align}
    \bm u(\bm r) = \sum_{\bm g} \bm u_{\bm g} e^{i \bm g \cdot \bm r}, \\
    h(\bm r) = \sum_{\bm g} h_{\bm g} e^{i \bm g \cdot \bm r},
\end{align}
where $\bm g$ are moir\'e reciprocal vectors and $h_{\bm g} = h_{-\bm g}^*$ and $\bm u_{\bm g} = \bm u_{-\bm g}^*$ are complex Fourier components. Using a Helmholtz decomposition for the in-plane components,
\begin{equation} \label{eq:ug}
    \bm u_{\bm g} = \frac{a}{L} \frac{\alpha_{\bm g} \hat z \times \bm g + \beta_{\bm g} \bm g}{ig^2},
\end{equation}
for $g = |\bm g| \neq 0$ and where $\alpha_{\bm g} = \alpha_{-\bm g}^*$ and $\beta_{\bm g} = \beta_{-\bm g}^*$ are complex numbers. Note that $\bm u_{\bm 0}$ is a constant relative shift between layers which does not affect the long-wavelength physics for small twists. Moreover, these coefficients are related to the divergence and curl:
\begin{align}
    \nabla \times \bm u & = \sum_{\bm g} i\bm g \times \bm u_{\bm g} e^{i \bm g \cdot \bm r} = \frac{\hat z a}{L} \sum_{\bm g} \alpha_{\bm g} e^{i\bm g \cdot \bm r}, \\
    \nabla \cdot \bm u & = \sum_{\bm g} i\bm g \cdot \bm u_{\bm g} e^{i \bm g \cdot \bm r} = \frac{a}{L} \sum_{\bm g} \beta_{\bm g} e^{i\bm g \cdot \bm r},
\end{align}
yielding the rotational and in-plane volumetric components of the displacement gradient. 

\begin{figure}
    \centering
    \includegraphics[width=.6\linewidth]{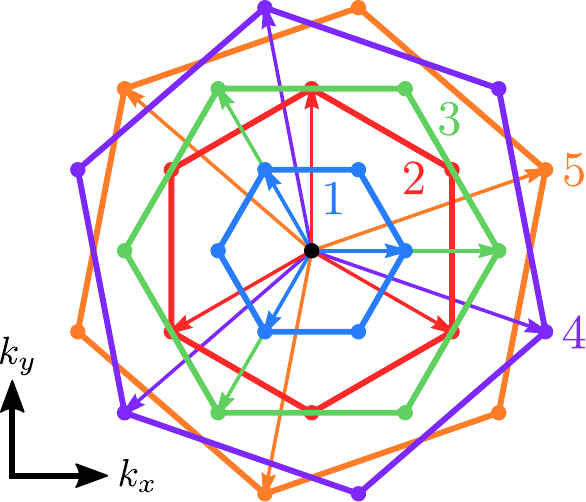}
    \caption{First five stars of reciprocal lattice vectors of the triangular lattice.}
    \label{fig:rstars}
\end{figure}

We now show how the displacement fields are constrained by symmetry. Under an in-plane symmetry $\mathcal S$, an in-plane vector field and a scalar function that preserve the symmetry have to transform as
\begin{align}
    \bm u(\mathcal S \bm r) & = \mathcal S u(\bm r), \\
    h(\mathcal S \bm r) & = h(\bm r), 
\end{align}
and similarly in reciprocal space,
\begin{align}
    \bm u_{\mathcal S\bm g} & = \mathcal S \bm u_{\bm g}, \\
    h_{\mathcal S\bm g} & = h_{\bm g}.
\end{align}
From Eq.\ \eqref{eq:ug} we then see that $\alpha_{\bm g}$ ($\beta_{\bm g}$) transforms as a pseudoscalar (scalar) under in-plane symmetries,
\begin{equation}
    \alpha_{\mathcal S \bm g} = \det(\mathcal S) \alpha_{\bm g}, \qquad \beta_{\mathcal S \bm g} = \beta_{\bm g}.
\end{equation}
On the other hand, for a symmetry that flips the layers, such as $\mathcal C_{2x}$ rotation symmetry, we have
\begin{align}
    \bm u(x,y) & = \begin{pmatrix} -1 & 0 \\ 0 & 1 \end{pmatrix} \bm u(x,-y), \\
    h(x,y) & = h(x,-y), 
\end{align}
since $\bm u_t(x,y) \mapsto \text{diag}(1,-1)\bm u_b(x,-y)$ and $h_t(x,y) \mapsto -h_b(x,-y)$ under $\mathcal C_{2x}$. To proceed, we first organize the reciprocal vectors of the triangular Bravais lattice in different stars, where each star contains six reciprocal vectors of equal magnitude and is invariant under $\mathcal C_{6z}$. For example, the first star is given by the six reciprocal vectors with $|\bm g| = 4\pi/\sqrt{3}L$. An illustration up to the fifth star is shown in Fig.\ \ref{fig:rstars}. 

We now discuss how the two generators of $D_6$ constrain the Fourier components. First, we see that $\mathcal C_{6z}$ rotation symmetry together with the reality of the displacements requires that each star is characterized by a single real Fourier coefficient which we call $\alpha_m$, $\beta_m$, and $h_m$ where $m = 1,2,\ldots$ labels the stars. Second, $\mathcal C_{2x}$ symmetry requires that $\beta_{g_x,g_y} = -\beta_{g_x,-g_y}$ while $\alpha_{g_x,g_y} = \alpha_{g_x,-g_y}$ and $h_{g_x,-g_y} = h_{g_x,-g_y}$. Hence, we find that the volumetric components of the first three stars vanish. For the degenerate fourth and fifth stars, we further find $\alpha_4 = \alpha_5$, $\beta_4 = -\beta_5$, and $h_4 = h_5$. Hence $D_6$ symmetry does not forbid but suppresses in-plane volumetric components.

In conclusion, up to the fifth star, there are four real rotational coefficients $\alpha_1$, $\alpha_2$, $\alpha_3$, and $\alpha_4 = \alpha_5$, as well as one in-plane volumetric coefficient $\beta_4 = -\beta_5$, and four real out-of plane coefficients $h_1$, $h_2$, $h_3$, and $h_4 = h_5$ (the constant term $h_0$ is always real). An overview of the symmetry-allowed parameters up to the fifth star is given in Table \ref{tab:D6} for both the homo and hetero displacements.
\begin{table}
    \label{tab:D6}
    \centering
    \begin{tabular}{c | c | c | c}
        \Xhline{1pt}
        $m$ & $\alpha_m$ & $\beta_m$ & $h_m$ \\
        \hline
        $1$ & $\mathds R$ & $0$ & $\mathds R$ \\ 
        $2$ & $\mathds R$ & $0$ & $\mathds R$ \\
        $3$ & $\mathds R$ & $0$ & $\mathds R$ \\
        $4$ & $\mathds R$ & $\mathds R$ & $\mathds R$ \\
        $5$ & $\alpha_4$ & $-\beta_4$ & $h_4$\\
        \Xhline{1pt}
    \end{tabular} \qquad
    \begin{tabular}{c | c | c | c}
        \Xhline{1pt}
        $m$ & $\overline \alpha_m$ & $\overline \beta_m$ & $\overline h_m$ \\
        \hline
        $1$ & $0$ & $\mathds R$ & $0$ \\ 
        $2$ & $0$ & $\mathds R$ & $0$ \\
        $3$ & $0$ & $\mathds R$ & $0$ \\
        $4$ & $\mathds R$ & $\mathds R$ & $\mathds R$ \\
        $5$ & $-\overline \alpha_4$ & $\overline \beta_4$ & $-\overline h_4$\\
        \Xhline{1pt}
    \end{tabular}
        \caption{Symmetry-allowed values for the in-plane and out-of-plane Fourier coefficients of the displacement fields in the presence of $D_6$ symmetry for the first five reciprocal stars.} 
\end{table}

\section{Beyond the first-star approximation} \label{app:fourier}

In the main text, we used the first-star approximation to find a minimal model for lattice relaxation in tBG, which works well for twist angles $\theta > 1.5^\circ$. Here, we address corrections due to more distant stars.

We start by taking taking the discrete Fourier transform of the \textsc{lammps} data. This yields the coefficients $\alpha_m$, $\beta_m$, and $h_m$ that were defined in Section \ref{app:symmetry}. Results for the rotational coefficients $\alpha_m$ of the relative displacement are shown up to the sixth star in Fig. (\textcolor{blue}{3}) of the main text. We do not show the volumetric in-plane components $\beta_m$, nor do we show the coefficients $\overline \alpha_m$ and $\overline \beta_m$ that are related to homostrain, as these are at least two orders of magnitude smaller than $\alpha_m$. The out-of-plane components are discussed below. We start by giving a simple theory for the coefficients $\alpha_1$, $\alpha_2$, and $\alpha_3$.

\subsection{Elastic theory} \label{app:elastic}

We consider a linear and isotropic elastic theory to model the intralayer strain and use the local-stacking approximation to model the interlayer van der Waals interactions. In particular, we minimize the energy functional $U = U_\text{elastic} + U_\text{vdW}$ given an \textit{ansatz} for the displacement fields that is motivated by symmetry and the \textsc{lammps} molecular dynamics simulations. The elastic and stacking-fault energy are given by \cite{nam2017lattice,carr2018relaxation,ceferino2023pseudomagnetic}
\begin{widetext}
\begin{align}
    U_\text{elastic}[ \bm u, \overline{\bm u}, h, \overline h] 
    & = \sum_{l=t,b} \int_\text{cell} d^2 \bm r \left\{ \frac{\lambda}{2} \left[ \text{tr} \left( \frac{u_l + u_l^t}{2} \right) \right]^2 + \mu \, \text{tr} \left( \frac{u_l + u_l^t}{2} \right)^2 + \frac{\kappa}{2} \left( \nabla^2 h_l \right)^2 \right\}, \label{eq:Helas} \\
    U_\text{vdW}[\bm u] 
    & = \sum_{\bm g} V_{\bm g} \int_\text{cell} d^2\bm r \, \exp \left\{ i \left[ \bm g \cdot \bm r + \frac{2L}{a} \, \hat z \times \bm g \cdot \bm u(\bm r) \right] \right\},
\end{align}
\end{widetext}
where $u_l$ is the displacement gradient for layer $l=t,b$, $V_{\bm g}$ are Fourier components of the stacking-fault energy, and $(L / a) \hat z \times \bm g$ is a reciprocal lattice vector of monolayer graphene. Assuming only a relative displacement between layers, we have $\bm u_{t/b} = \pm \bm u$ and $h_{t/b} = \pm h$ such that
\begin{equation}
    \left( u_{t/b} \right)_{ij} = \pm \frac{\partial u_i}{\partial r_j} + \frac{1}{2}  \frac{\partial h}{\partial r_i} \frac{\partial h}{\partial r_j},
\end{equation}
respectively. We see that the minus sign from heterostrain only appears for the in-plane components since the out-of-plane displacements contribute at second order. This actually leads to a cancellation of cross terms such that the in-plane motion becomes decoupled from the out-of-plane motion. Immediately one finds a trivial solution for the out-of-plane displacement field $h = \text{constant}$. Note that this theory lacks information on the preferred height profile of the twisted bilayer system. In principle, this would modify $U_\text{vdW}$ and couple the in-plane and out-of-plane displacements \cite{mchugh2023_moire}. However, we assume here that the height profile follows the optimal value given by the local stacking, which agrees well with the \textsc{lammps} simulations. Likewise, if one completely neglects out-of-plane displacements, then the homo and hetero displacements are decoupled: $U_\text{elastic}[\bm u, \overline{\bm u}] = 2U_\text{elastic}[\bm u] + 2U_\text{elastic}[\overline{\bm u}]$ such that $\overline{\bm u}$ vanishes in equilibrium.

We now further assume that the displacements are small such that $|\bm u| \ll a$. Then we can expand the van der Waals energy density in powers of $|\bm b_i \cdot \bm u|$. This yields
\begin{widetext}
\begin{align}
    U_\text{vdW} & = \sum_{\bm g} V_{\bm g} \int d^2 \bm r \, e^{i\bm g \cdot \bm r} \left\{ 1 + \frac{2iL}{a} \, \hat z \times \bm g \cdot \bm u(\bm r) - \frac{1}{2} \left[ \frac{2L}{a} \, \hat z \times \bm g \cdot \bm u(\bm r) \right]^2 + \cdots \right\} \\
    & = A \sum_{\bm g} V_{\bm g} \left( \delta_{\bm g, \bm 0} - 2 \alpha_{-\bm g} - \frac{2L^2}{a^2} \sum_{\bm g' \neq \bm g} \left( \hat z \times \bm g \cdot \bm u_{-\bm g'} \right) \left( \hat z \times \bm g \cdot \bm u_{\bm g' - \bm g} \right) + \cdots \right) \\
    & = A V_{\bm 0} - 2 A \sum_{\bm g} V_{\bm g} \alpha_{-\bm g} + \mathcal O(|\bm u|^2/a^2),
\end{align}
\end{widetext}
where $A = \sqrt{3} L^2/2$ is the moir\'e cell area and we used Eq.\ \eqref{eq:ug}. Therefore, in lowest order the stacking-fault energy only contains contributions from the rotational components of the in-plane displacement field. This feature is universal for \textit{twisted} moir\'e materials and can be traced back to the local disregistry of the atoms,
\begin{equation} \label{eq:disreg}
    \bm d(\bm r) = 2 \sin(\theta/2) \, \hat z \times \bm r + 2 \bm u(\bm r).
\end{equation}

To proceed, we expand the in-plane displacement field up to the third star of reciprocal vectors. We also know from our symmetry analysis that $D_6$ forbids $\nabla \cdot \bm u$ up to the third star, giving three rotational coefficients $\alpha_1$, $\alpha_2$, and $\alpha_3$ to be determined. Under these approximations, that are justified by the molecular dynamics simulation,
\begin{align}
    U_\text{elastic} & = \mu \int d^2 \bm r \left[ \left( u_{xx} - u_{yy} \right)^2 + \left( u_{xy} + u_{yx} \right)^2 \right] \\
    & = \frac{6 A a^2 \mu}{L^2} \left( \alpha_1^2 + \alpha_2^2 + \alpha_3^2 \right).
\end{align}
Next, we expand the integrand of $U_\text{vdW}$. We only consider the first three stars of $V_{\bm g}$ \cite{carr2018relaxation}. Moreover, since the stacking-fault energy for tBG and homobilayer tTMDs near parallel stacking has $D_6$ symmetry, the coefficients $V_1$, $V_2$, and $V_3$ need to be real. Expanding $\alpha_1$, $\alpha_2$, and $\alpha_3$ up to second order yields
\begin{equation}
    \begin{aligned}
        U_\text{vdW} & \simeq 6 A \Big\{ V_1 \alpha_1 \left[ \alpha_1 - 2 \left( 1 + \alpha_2 + \alpha_3 \right) \right] \\
        & + V_2 \left[ 9 \alpha_1^2 + \alpha_2 \left( \alpha_2 - 2 \right) \right] \\
        & + V_3 \left[ 8 \alpha_1 \left( \alpha_1 + \alpha_2 \right) + \alpha_3 \left( \alpha_3 - 2 \right) \right] \Big\} .
    \end{aligned}
\end{equation}
Minimizing the energy $U$ with respect to $\alpha_1$, $\alpha_2$, and $\alpha_3$ then gives 
\begin{align}
    \alpha_1 & \simeq \frac{c_1}{4 \sin^2 \tfrac{\theta}{2}} - \frac{c_1^2}{16 \sin^4 \tfrac{\theta}{2}}, \label{eq:Salpha1} \\
    \alpha_{2,3} & \simeq \frac{c_{2,3}}{4 \sin^2 \tfrac{\theta}{2}} + \frac{c_1^2}{16 \sin^4 \tfrac{\theta}{2}}, \label{eq:Salpha23}
\end{align}
with $c_m = V_m / \mu$ dimensionless materials constants that can be determined from a local-stacking approximation and DFT calculations \cite{carr2018relaxation} or from molecular dynamics simulations. In the main text, we take an alternative approach and fit the coefficients $\alpha_m$ to a series expansion in $(L/a)^n$. The coefficients of this series (taken up to fourth order) are given in Table \ref{tab:params}.
\begin{table*}
    \centering
    \begin{tabular}{| c |c | c | c | c |}
        \Xhline{1pt}
        \backslashbox{$i$}{$n$} & $1$ & $2$ & $3$ & $4$ \\ \hline
        $1$ & $6.87 \times 10^{-5}$ & $-5.82 \times 10^{-9}$ & $-7.13 \times 10^{-13}$ & $1.33 \times 10^{-16}$ \\ \hline
        $2$ & $-3.10 \times 10^{-5}$ & $5.44 \times 10^{-9}$ & $-1.23 \times 10^{-12}$ & $9.60 \times 10^{-17}$ \\ \hline
        $3$ & $-1.34 \times 10^{-6}$ & $4.53 \times 10^{-9}$ & $-7.50 \times 10^{-13}$ & $4.89 \times 10^{-17}$ \\ \Xhline{1pt}
    \end{tabular} \\ \vspace{2mm}
    \begin{tabular}{| c |c | c | c | c |}
        \Xhline{1pt}
        \backslashbox{$i$}{$n$} & $1$ & $2$ & $3$ & $4$ \\ \hline
        $1$ &$4.28 \times 10^{-4}$ & $-2.96 \times 10^{-7}$ & $8.79 \times 10^{-11}$ & $-6.51 \times 10^{-15}$ \\ \hline
        $2$ &$-4.51 \times 10^{-5}$ & $1.81 \times 10^{-7}$ & $-1.51 \times 10^{-10}$ & $3.92 \times 10^{-14}$ \\ \hline
        $3$ & $-3.68 \times 10^{-5}$ & $2.09 \times 10^{-7}$ & $-1.46 \times 10^{-10}$ & $3.37 \times 10^{-14}$ \\ \hline
        $6$ & $-5.01 \times 10^{-6}$ & $3.61 \times 10^{-9}$ & $1.94 \times 10^{-11}$ & $-7.82 \times 10^{-15}$ \\
        \Xhline{1pt}
    \end{tabular}
    \label{tab:params}
    \caption{Coefficients $A_{in}$ for tBG (top table) and tWSe$_2$ (bottom table) corresponding to the numerical fits from Fig. (\textcolor{blue}{3}) of the main text to the series in Eq.(\textcolor{blue}{6}). Here $i$ indexes the star and $n$ the power of $(L/a)^2$. Here we fit using least squares using data down to $0.86^\circ$ for tBG and $1.5^\circ$ for tWSe$_2$ using the displacements of W atoms.} 
\end{table*}

\subsection{Out-of-plane displacements} \label{app:outofplane}

The discussion in the main text was focused on the in-plane displacement field. In this section, we present a simple theory for the out-of-plane displacement field.

Consider an untwisted bilayer with a constant relative shift $\bm d$ between layers. The interlayer distance, defined as $2h(\bm d)$, is an even periodic function of $\bm d$ with periods given by the primitive lattice vectors $\bm a_1$ and $\bm a_2$ of the monolayer. Moreover, it is invariant under $\bm d \rightarrow -\bm d$ and threefold rotations. Hence, we can approximate it as \cite{koshino2018_maximally}
\begin{equation}
    h(\bm d) = \tilde h_0 + 2 \tilde h_1 \sum_{i=1}^3 \cos(\bm b_i \cdot \bm d),
\end{equation}
where the sum runs over graphene reciprocal lattice vectors of the first star, related by $\mathcal C_{3z}$. Here
\begin{align}
    \tilde h_0 & = \frac{h_\text{AA} + 2 h_\text{AB}}{6}, \label{eq:h0} \\
    \tilde h_1 & = \frac{h_\text{AA} - h_\text{AB}}{18}, \label{eq:h1}
\end{align}
where $h_\text{AA}$ and $h_\text{AB}$ is the interlayer distance for AA [$\bm d = \left( 0, 0 \right)$] and AB [$\bm d = \left( 0, a/\sqrt{3} \, \right)$] stacking, respectively.

For the twisted bilayer, we can use a local stacking approximation where we view the local lattice structure approximately in terms of an untwisted bilayer with local disregistry given by Eq.\ \eqref{eq:disreg}. This approximation yields
\begin{equation}
    h(\bm r) = \tilde h_0 + 2 \tilde h_1 \sum_{i=1}^3 \cos \left[ \bm g_i \cdot \bm r + 2 \bm b_i \cdot \bm u(\bm r) \right].
\end{equation}
Expanding in lowest order of $|\bm b_i \cdot \bm u(\bm r)|$ modifies the Fourier coefficients of the zeroth and first star as
\begin{align}
    h_0 & = \tilde h_0 - 12 \tilde h_1 \alpha_1 = \tilde h_0 - \frac{3 \tilde h_1 c_1}{\sin^2 \tfrac{\theta}{2}}, \\
    h_1 & = \tilde h_1 \left( 1 - 2 \alpha_1 \right) = \tilde h_1 \left( 1 - \frac{c_1}{2\sin^2 \tfrac{\theta}{2}} \right),
\end{align}
and gives rise to second and third star coefficients
\begin{equation}
    h_2 = h_3 = 2 \tilde h_1 \alpha_1 = \frac{\tilde h_1 c_1}{2\sin^2 \tfrac{\theta}{2}}.
\end{equation}
Hence the scaling laws for out-of-plane displacements are approximately inherited from the in-plane displacements. 
Second-order corrections by taking into account terms of order $\alpha_1^2$, $\alpha_2$, and $\alpha_3$ give rise to components in $h(\bm r)$ up to the sixth star. We list them here for completeness:
\begin{align}
    h_0 & = \tilde h_0 - 6 \tilde h_1 \alpha_1 \left( 2 - \alpha_1 \right), \\
    h_1 & = \tilde h_1 \left( 1 - 2 \alpha_1 + 4 \alpha_1^2 + 2 \alpha_2 + \alpha_3 \right), \\
    h_2 & = \tilde h_1 \left[ \left( 2 - 3 \alpha_1 \right) \alpha_1 - \alpha_3 \right], \\
    h_3 & = \tilde h_1 \left( 2 - \alpha_1 \right) \alpha_1, \\
    h_4 = h_5 & = \frac{\tilde h_1}{2} \left( 5 \alpha_1^2 + 2 \alpha_2 + \alpha_3 \right), \\
    h_6 & = \tilde h_1 \left( 2 \alpha_1^2 + \alpha_3 \right).
\end{align}

\begin{figure}
    \centering
    \includegraphics[width=\linewidth]{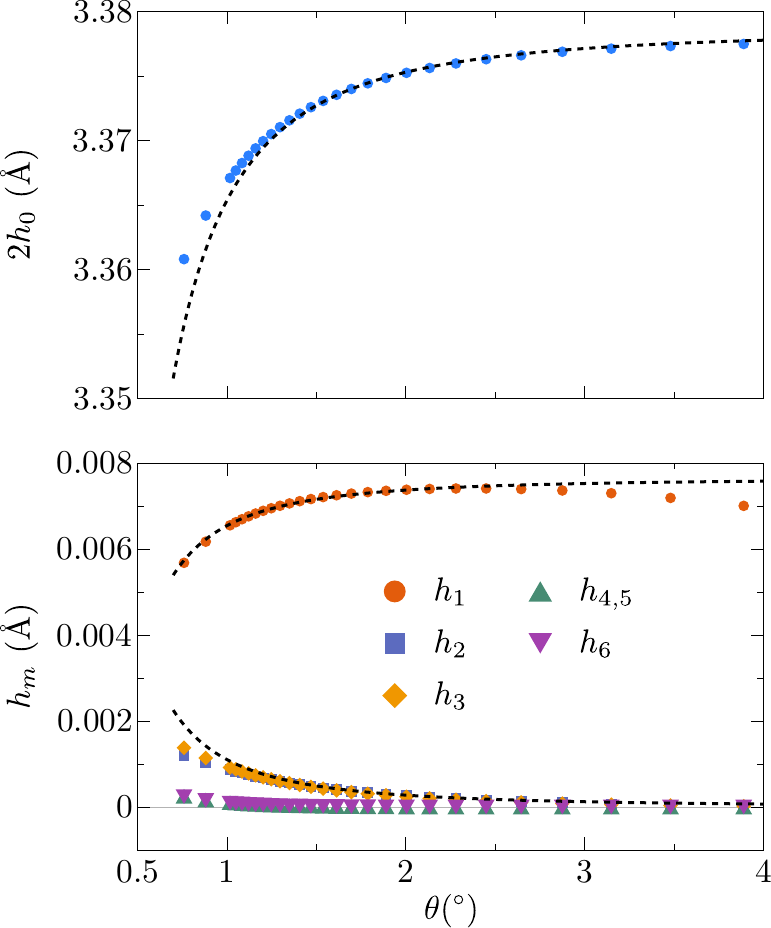}
    \caption{Corrugation in tBG. (a) Average interlayer distance $2h_0$ versus twist angle. (b) Fourier components of the relative out-of-plane displacement field $h(\bm r)$. Dots are calculated with \textsc{lammps} for DRIP1 interlayer potential and dashed lines are fits to the local-stacking approximation in lowest order.}
    \label{fig:tbg_fourier_h}
\end{figure}

This local stacking approximation does not work as well for the DRIP2 force field used earlier.  Instead, we fit these analytical results for the out-of-plane relaxation to \textsc{lammps} data obtained using the DRIP1 potential \cite{JeilFF}.  As before, the single parameter of the analytical theory is determined just from the in-plane relaxation data where we find $c_1 = 2.2 \times 10^{-5}$.  One consequence of the smaller value of $c_1$ for DRIP1 compared to DRIP2 is that the scaling theory for relaxation works even better in this case compared to the data shown in the main text. Fitting the lowest order expressions above then yields $\tilde h_0 \approx 1.6893 \, \text{\AA}$ and $\tilde h_1 \approx 7.65 \times 10^{-3} \, \text{\AA}$. Using Eqs.\ \eqref{eq:h0} and \eqref{eq:h1} we obtain $h_\text{AB} \approx 3.33 \, \text{\AA}$ and $h_\text{AA} \approx 3.47 \, \text{\AA}$. This agrees with the interlayer distance at the AB and AA stacking centers for small twist angles, respectively. The resulting fits are shown in Fig.\ \ref{fig:tbg_fourier_h}. 

\begin{figure}
    \centering
    \includegraphics[width=\linewidth]{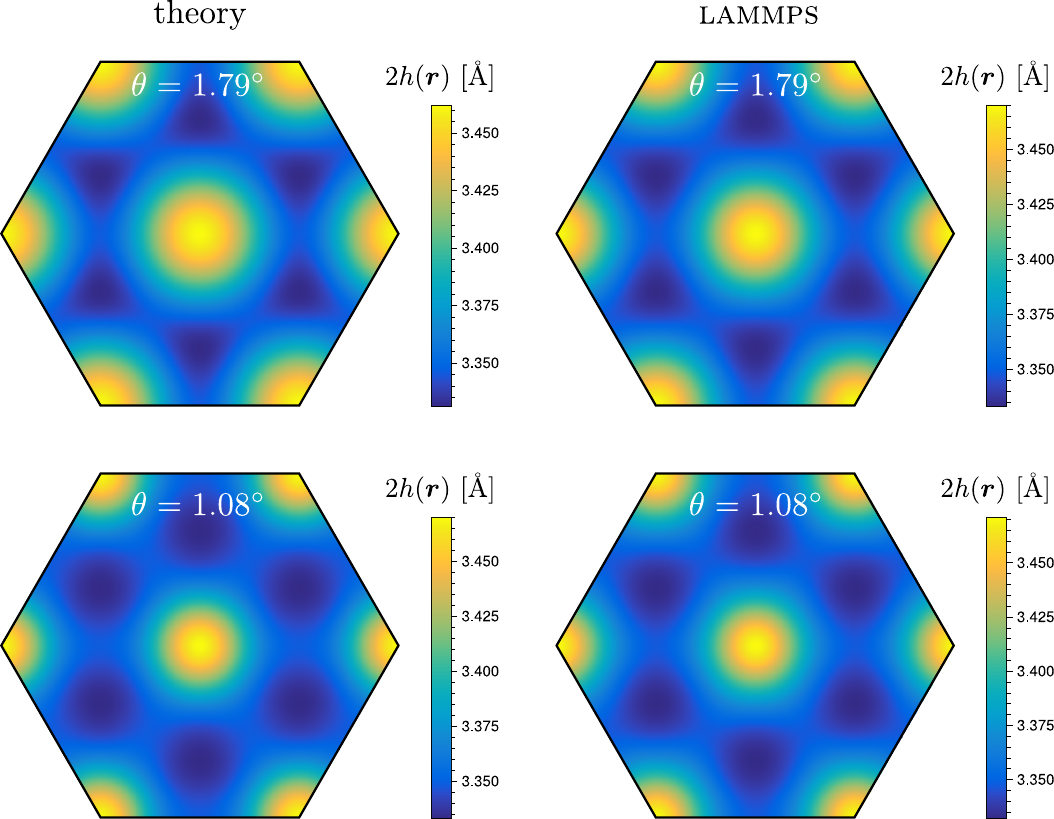}
    \caption{Interlayer distance for tBG comparing theory (left) and \textsc{lammps} for DRIP1 interlayer potential (right). Theory is based on the local-stacking approximation in lowest order including up to three reciprocal stars.}
    \label{fig:tbg_ild}
\end{figure}

In Fig.\ \ref{fig:tbg_ild} we show the interlayer distance for tBG for two different twist angles comparing theory and \textsc{lammps} emphasising that the theory only contains the parameter $c_1$ that is fully determined by the in-plane relaxation using Eq. (\textcolor{blue}{4}). Finally, we show the interlayer distance for tWSe$_2$ calculated from \textsc{lammps} for three representative twist angles in Fig.\ref{fig:tbg_ild}. We find that $\theta = 5^\circ$ and $\theta = 3.48^\circ$ converges for $1$ and $3$ stars, respectively, while for $\theta = 2^\circ$ we require $6$ stars.
\begin{figure*}
    \centering
    \includegraphics[width=\linewidth]{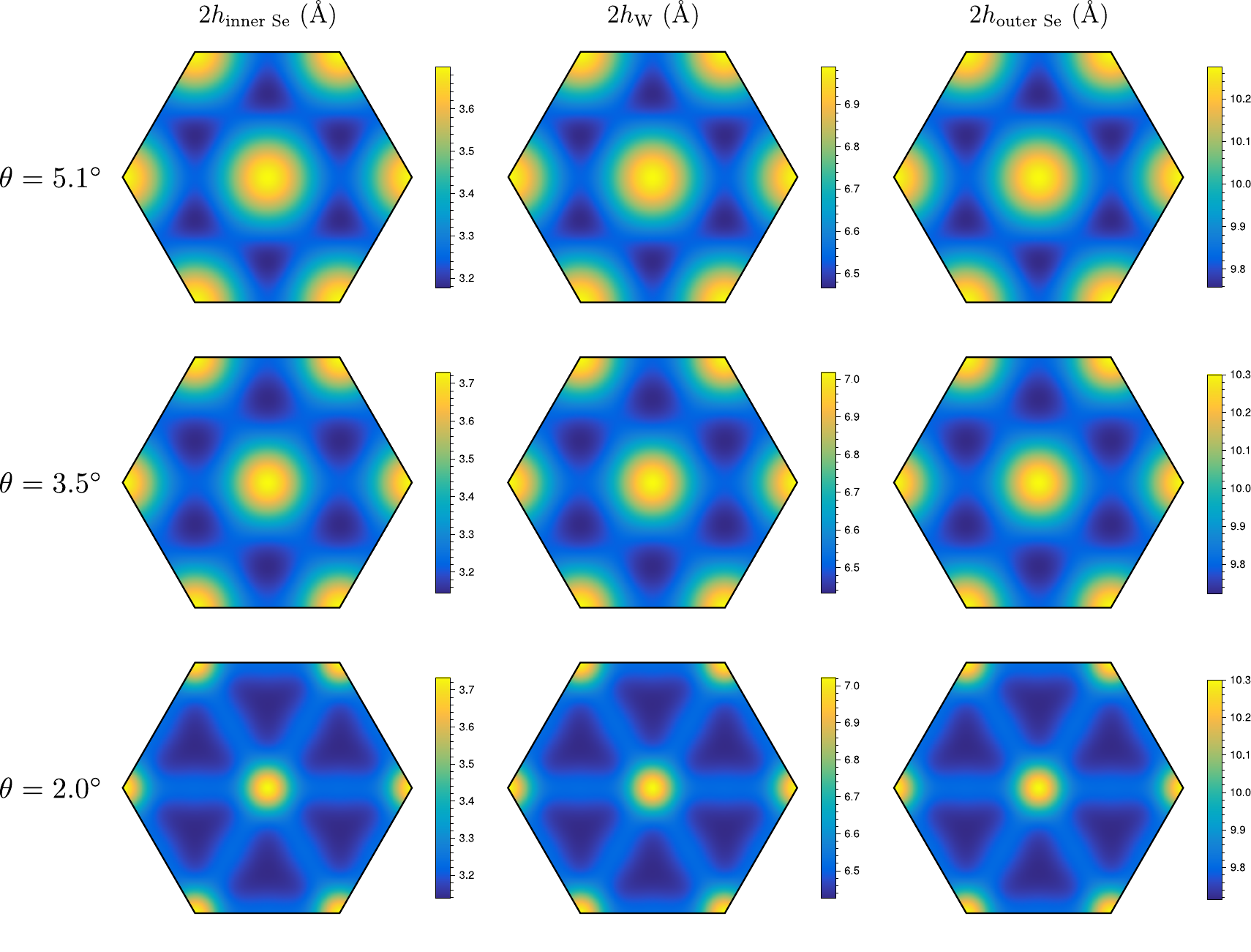}
    \caption{Interlayer distance for tWSe$_2$ calculated with \textsc{lammps} by discrete Fourier transform. We show both the distance between inner and outer chalcogen atoms and metal atoms.}
    \label{fig:twse2_ild}
\end{figure*}

\section{Continuum model from symmetry} \label{app:continuum}

In this section, we derive the continuum model from symmetry. For small twist angles, the valleys are effectively decoupled, and the low-energy Hamiltonian is given by a Bistritzer-MacDonald (B-M) model \cite{bistritzer2011moire}
\begin{equation}
    H = \sum_{\nu = \pm 1} \int d^2 \bm r \, \psi_\nu^\dag(\bm r)
    \begin{bmatrix}
    H_t & T(\bm r) \\ T^\dag(\bm r) & H_b
    \end{bmatrix} \psi_\nu(\bm r),
\end{equation}
with valley index $\nu$ and where $\psi_\nu(\bm r) = \left[ \psi_{\nu t}(\bm r), \psi_{\nu b}(\bm r) \right]^t$ are four-component field operators. In the following, we choose a coordinate system where the $x$ axis of the lies along the zigzag direction of the graphene with primitive lattice vectors $\bm a_{1,2} = a ( \pm 1/2, \sqrt{3}/2)$ where $a = 0.246 \, \text{nm}$ is the graphene lattice constant.

\subsection{Intralayer Hamiltonian}

The intralayer Hamiltonian can be written as
\begin{equation} \label{eq:intraH}
    H_{t/b} = \hbar v_F \left[ R_{\mp\theta/2} \left( - i \nabla - \nu \bm K_{t/b} \right) \right] \cdot \left( \nu \sigma_x , \sigma_y \right),
\end{equation}
where $v_F = 1.05 \times 10^6 \, \text{m/s}$ is  the monolayer graphene Fermi velocity and $\bm K_{t/b} = k_\theta \left( \sqrt{3}/2, \pm 1/2 \right)$ with $k_\theta = 4 \pi / 3 L$ and $L = a / 2 \sin \left( \theta / 2 \right)$ the moir\'e period. Here we have chosen the momentum origin at the center of the moir\'e Brillouin zone and $\sigma_{x,y}$ are Pauli matrices that act in sublattice space. We do not take into account a symmetry-allowed sublattice potential due to different atomic environments for $A$ and $B$ atoms \cite{ceferino2023pseudomagnetic,garcia2021_full}. 

The intralayer Hamiltonian can be simplified by a gauge transformation
\begin{equation}
    \psi_{\nu l}(\bm r) \mapsto e^{i\nu \bm K_l \cdot \bm r} \psi_{\nu l}(\bm r),
\end{equation}
yielding
\begin{equation}
    H_{t/b} = -i \hbar v_F R_{\mp\theta/2} \nabla \cdot \left( \nu \sigma_x , \sigma_y \right).
\end{equation}
For this gauge choice, the continuum model can be diagonalized by Fourier transform ($l = t, b$)
\begin{equation}
    \psi_{\nu l}(\bm r) = \sum_{\bm k \in \text{MBZ}} \sum_{\bm g} e^{i \left( \bm k - \nu \bm K_l - \bm g \right) \cdot \bm r} c_{\nu l}(\bm k - \bm g).
\end{equation}

\subsection{Moir\'e coupling} \label{moirecoupling}

We now consider the interlayer moir\'e potential in the lowest harmonic. Here we only consider a local moir\'e potential. Symmetry constraints on nonlocal contributions are discussed in Ref.\ \cite{kang2023}. In the original gauge, the lowest moir\'e harmonic becomes
\begin{equation}
    T(\bm r) = T_1 + T_2 e^{-i \nu \bm g_1 \cdot \bm r} + T_3 e^{-i \nu \left( \bm g_1 + \bm g_2 \right) \cdot \bm r},
\end{equation}
with $\bm g_1 = -\sqrt{3} k_\theta \left( 1/2,  \sqrt{3}/2 \right)$ and $\bm g_2 = \sqrt{3} k_\theta \left( 1, 0 \right)$ reciprocal lattice vectors of the moir\'e lattice related by $\mathcal C_{3z}$ symmetry. In the new gauge, we have
\begin{equation} \label{eq:T}
    T(\bm r) \mapsto T(\bm r) e^{i \nu \left( \bm K_b - \bm K_t \right) \cdot \bm r} = \sum_{i=1}^3 T_j e^{i \nu \bm q_j \cdot \bm r},
\end{equation}
with $\bm q_1 = k_\theta \left( 0, -1 \right)$ and $\bm q_{2,3} = k_\theta \left( \pm \sqrt{3}/2, 1/2 \right)$. We work in this gauge for the remainder of this section.

The complex $2 \times 2$ matrices $T_j$ ($j=1,2,3$) are constrained by the (emergent) symmetries of the moir\'e lattice that preserve the valley index. The valley-preserving symmetries form the dichromatic group 6\textquotesingle2\textquotesingle2 also denoted as $D_6(D_3) = D_3 + (D_6 \setminus D_3 ) \mathcal T$ \cite{dresselhaus_group_2007} and generated by:
\begin{itemize}
\item $\mathcal C_{2z} \mathcal T$: composition of spinless time reversal $\mathcal T$ with a $\pi$ rotation about the $z$ axis. This operation leaves the layers invariant but exchanges the sublattices;
\item $\mathcal C_{3z}$: rotation by $\pm 2\pi/3$ about $z$. Leaves the sublattices invariant up to a phase factor from rotating the pseudospin;
\item $\mathcal C_{2x}$: rotation by $\pi$ about the $x$ axis. Exchanges both layers and sublattices;
\end{itemize}
as illustrated in Fig.\ \ref{fig:tbgD6}(b).

\subsubsection*{$\mathcal C_{2z} \mathcal T$ symmetry}

First, we consider $\mathcal C_{2z} \mathcal T$ symmetry. Its action on the field operators is represented by
\begin{equation}
    \left( \mathcal C_{2z} \mathcal T \right) \psi_\nu(\bm r) \left( \mathcal C_{2z} \mathcal T \right)^{-1} = \tau_0 \sigma_x \psi_\nu(-\bm r),
\end{equation}
where $\tau$ matrices act in layer space. From $[ H , \mathcal C_{2z} \mathcal T ] = 0$ and $\mathcal T i \mathcal T^{-1} = -i$, we find that the intralayer Hamiltonian is invariant as expected, while the interlayer coupling needs to satisfy
\begin{equation}
    T(\bm r) = \sigma_x T(-\bm r)^* \sigma_x.
\end{equation}
For the first moir\'e harmonic, taking $T_j$ as a general $2 \times 2$ complex matrix:
\begin{equation}
    T_j = \sigma_x T_j^* \sigma_x = a_j \sigma_0 + b_j \sigma_x + c_j \sigma_y + i d_j \sigma_z,
\end{equation}
with $\{a_j,b_j,c_j,d_j\}$ real constants (12 in total). Note that this result holds for any moir\'e harmonic. For example, the second moir\'e harmonic lies at distance $2k_\theta$ from the principal Dirac point of a given layer.

\subsubsection*{$\mathcal C_{3z}$ symmetry}

While $\mathcal C_{2z} \mathcal T$ constrains each matrix individually, we now show that $\mathcal C_{3z}$ rotation symmetry gives a relation between different $T_j$ matrices. The action on the field operators is given by
\begin{equation}
    \mathcal C_{3z} \psi_\nu(\bm r) \mathcal C_{3z}^{-1} = \tau_0 e^{i \nu \pi \sigma_z /3} \psi_\nu(\mathcal C_{3z} \bm r).
\end{equation}
One way to understand the sublattice rotation is that it is required to keep $-i \nabla \cdot \left( \nu \sigma_x , \sigma_y \right)$ invariant. Indeed,
\begin{widetext}
\begin{equation}
    -i \nabla \cdot \left( \nu \sigma_x , \sigma_y \right) \mapsto -i R_{-2\pi/3} \nabla \cdot \left[ e^{-i \nu \pi \sigma_z /3} \left( \nu \sigma_x , \sigma_y \right) e^{i \nu \pi \sigma_z /3} \right] = - i  \nabla \cdot \left( \nu \sigma_x , \sigma_y \right).
\end{equation}
\end{widetext}

For the interlayer term, we obtain a relation between the three matrices of each moir\'e harmonic, reducing the number of real parameters from $12$ to $4$. Explicitly,
\begin{align}
    T(\bm r) & = e^{-i\nu \pi \sigma_z/3} T(\mathcal C_{3z}^{-1} \bm r) e^{i\nu \pi \sigma_z/3} \\
    & = \sum_{j=1}^3 e^{-i\nu\pi \sigma_z/3} T_j e^{i\nu\pi \sigma_z/3} e^{i\nu \pi \sigma_z/3} e^{i \bm q_{j+1} \cdot \bm r},
\end{align}
where $\bm q_4 = \bm q_1$.
We thus require
\begin{equation}
    T_{j+1} = e^{-i\nu\pi \sigma_z/3} T_j e^{i\nu\pi \sigma_z/3}.
\end{equation}
where $j=1,2,3$ is defined cyclically. Hence,
\begin{align}
    T_1 & = a \sigma_0 + i d \sigma_z + b \sigma_x + c \sigma_y, \\
    T_2 & = a \sigma_0 + i d \sigma_z + e^{-\frac{i\nu\pi}{3} \sigma_z} \left( b \sigma_x + c \sigma_y \right) e^{+\frac{i\nu\pi}{3} \sigma_z}, \\
    T_3 & = a \sigma_0 + i d \sigma_z + e^{+\frac{i\nu\pi}{3} \sigma_z} \left( b \sigma_x + c \sigma_y \right) e^{-\frac{i\nu\pi}{3} \sigma_z}.
\end{align}

\subsubsection*{$\mathcal C_{2x}$ symmetry}

Lastly we consider $\mathcal C_{2x}$ symmetry which exchanges layers and sublattices but leaves the valley invariant. Its action is given by
\begin{equation}
    \mathcal C_{2x} \psi_\nu(x,y) \mathcal C_{2x}^{-1} = \tau_x \sigma_x \psi_\nu(x,-y),
\end{equation}
which leaves the intralayer Hamiltonian invariant, but gives a constraint on the interlayer coupling. Restricting to the first moir\'e harmonic, we find
\begin{equation}
    T(x,y) = \sigma_x T^\dag(x,-y) \sigma_x,
\end{equation}
or
\begin{equation}
    T_1 = \sigma_x T_1^\dag  \sigma_x, \qquad T_2 = \sigma_x T_3^\dag \sigma_x,
\end{equation}
which sets $c=0$.

Finally, we obtain
\begin{align}
    T_j & = w_1 e^{i \phi \sigma_z} + w_2 \left[ \sigma_x \cos \frac{2\pi j}{3} + \nu \sigma_y \sin \frac{2\pi j}{3} \right] \\
    & = \begin{pmatrix} w_1 e^{i\phi} & w_2 e^{-i\nu2\pi j/3} \\ w_2 e^{i\nu2\pi j/3} & w_1 e^{-i\phi} \end{pmatrix},
    \label{phisymmetry}
\end{align}
where $a = w_1 \cos \phi$, $d = w_1 \sin \phi$, and $w_2 = b$ are real parameters. The magnetic point group 6\textquotesingle2\textquotesingle2 thus yields three real parameters for the first moir\'e harmonic.
\begin{figure*}
    \centering
    \includegraphics[width=.7\linewidth]{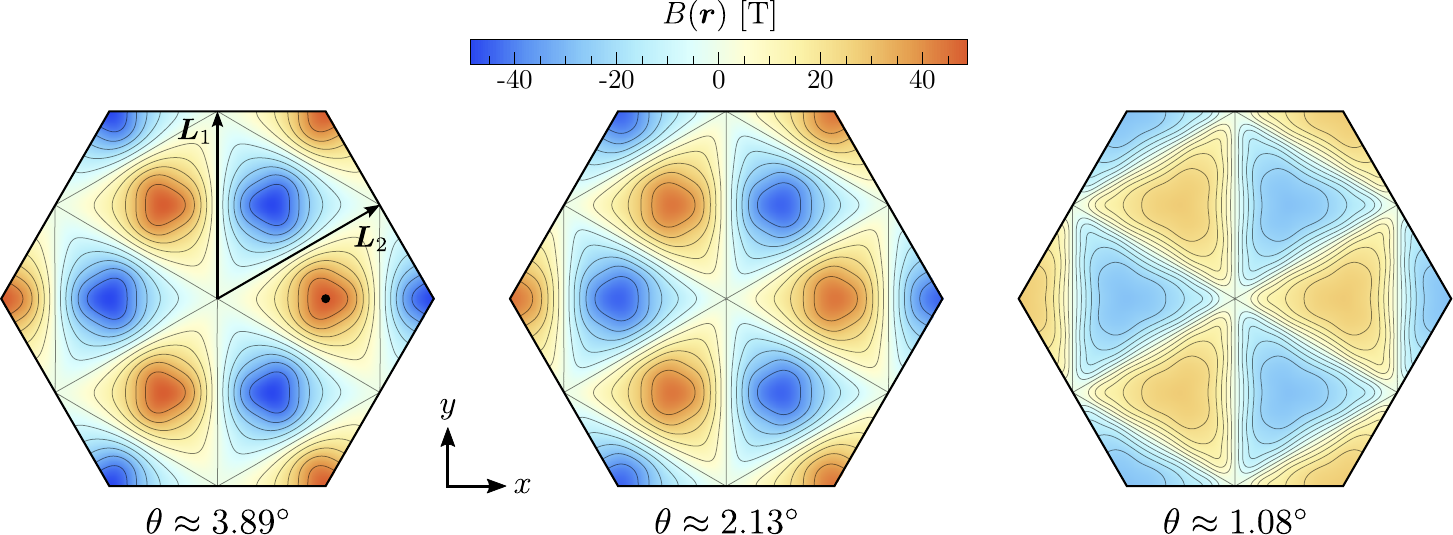}
    \caption{Pseudo magnetic field from heterostrain calculated from the \textsc{lammps} simulation for tBG for three twist angles.}
    \label{fig:pmf}
\end{figure*}

\subsection{Discussion}

A similar symmetry analysis can be found in Ref.\ \cite{balents_general_2019} but the phase $\phi$ of the AA coupling was not considered in this work. At a first glance, one might think that this phase can be removed by a unitary. Indeed, consider
\begin{equation}
    \begin{pmatrix}
    \psi_{\nu t} \\ \psi_{\nu b}
    \end{pmatrix} \mapsto 
    e^{i \phi \tau_z \sigma_z / 2} \begin{pmatrix}
    \psi_{\nu t} \\ \psi_{\nu b}
    \end{pmatrix},
\end{equation}
such that
\begin{equation}
    \psi_{\nu t}^\dag T \psi_{\nu b} \mapsto \psi_{\nu t}^\dag e^{-i\phi \sigma_z/2} T e^{-i\phi \sigma_z/2} \psi_{\nu b},
\end{equation}
where
\begin{equation}
    \begin{aligned}
        & e^{-i\phi \sigma_z/2} T_j e^{-i\phi \sigma_z/2} \\
        & = w_1 \sigma_0 + w_2 \left[ \sigma_x \cos \frac{2\pi j}{3} + \nu \sigma_y \sin \frac{2\pi j}{3} \right].
    \end{aligned}
\end{equation}
However, this unitary affects the intralayer part of the Hamiltonian. Namely,
\begin{widetext}
\begin{align}
    -i R_{\mp \theta / 2} \nabla \cdot \left( \nu \sigma_x , \sigma_y \right) & \mapsto -i R_{\mp \theta / 2} \nabla \cdot \left[ e^{\mp i \phi \sigma_z / 2} \left( \nu \sigma_x , \sigma_y \right) e^{\pm i \phi \sigma_z / 2} \right] = -i R_{\mp \left( \theta/2 - \phi \right)} \nabla \cdot \left( \nu \sigma_x , \sigma_y \right).
\end{align}
\end{widetext}
This shows that the phase $\phi$ of the AA interlayer moir\'e coupling is equivalent to a rotation of the sublattice pseudospin in the intralayer Hamiltonian. Hence, we can think of $\phi$ as having a rigid contribution which is negligible in the limit of small twist angles.

\subsection{Pseudo gauge fields} \label{app:pseudo}

In the long-wavelength limit, it is known that strain fields couple to the low-energy electronic degrees of freedom of graphene as effective gauge fields \cite{vozmediano_gauge_2010,amorim_novel_2016}. In particular, shear strain gives rise to a pseudo vector potential $\pm \bm A_l(\bm r)$ at valley $\bm K_\pm = \left( \pm 4\pi/3a, 0 \right)$ for $l = t, b$. For twisted bilayer graphene, we only consider the pseudo gauge field arising from heterostrain. The intralayer Hamiltonian in Eq.\ \eqref{eq:intraH} is then modified by replacing $-i\nabla \rightarrow -i \nabla + \nu e \bm A_l(\bm r) / \hbar$.

In a coordinate system where the $x$ axis lies along the zigzag direction of the original untwisted graphene, the pseudo vector potential is given by \cite{vozmediano_gauge_2010}
\begin{equation}
    \bm A_t(\bm r) = \bm A(\bm r) = \frac{\sqrt{3}\hbar \beta}{2ea} \begin{pmatrix} u_{tyy} - u_{txx} \\ u_{txy} + u_{tyx} \end{pmatrix},
\end{equation}
where $\beta = \tfrac{a}{\sqrt{3}t_0} \left| \tfrac{\partial t}{\partial r} \right|_{nn} \approx 2$ is the electron Gr\"uneisen parameter and $-e$ is the electron charge. Here we defined the displacement gradient ($l=t,b$)
\begin{equation}
    u_{lij} = \frac{\partial u_{li}}{\partial r_j} + \frac{1}{2} \frac{\partial h_l}{\partial r_i} \frac{\partial h_l}{\partial r_j},
\end{equation}
which is obtained by considering the change in length after a deformation between two points with initial infinitesimal and in-plane separation $dr_i$ \cite{landau_theory_1986}. Explicitly, we have $\left[ (dr_i + du_i)^2 + dh^2 \right ] - dr_i^2 \equiv 2u_{ij} dr_i dr_j$ where the left-hand side is evaluated up to lowest order in the displacements. To preserve $\mathcal C_{2x}$ symmetry, we further have
\begin{equation} \label{eq:Ab}
    \bm A_b(x, y) = \begin{pmatrix} 1 & 0 \\ 0 & -1 \end{pmatrix} \bm A_t(x, -y).
\end{equation}
This can be derived from the continuum model or from the transformation of the displacement gradient: $\left[ u_b(x,y) \right]_{ij} = \left( \sigma_z \right)_{ik} \left[ u_t (x,-y) \right]_{kl} \left( \sigma_z \right)_{lj}$. The remaining symmetries of the 6\textquotesingle2\textquotesingle2 magnetic point group of a single valley, yield, up to a gauge transformation,
\begin{equation}
    \bm A(\bm r) = \bm A(-\bm r) = \mathcal C_{3z} \bm A(\mathcal C_{3z}^{-1} \bm r),
\end{equation}
which, together with Eq.\ \eqref{eq:Ab}, implies
\begin{align}
    B(\bm r) & = -B(-\bm r) = B(\mathcal C_{3z}^{-1} \bm r), \\
    B_b(x,y) & = -B_t(x,-y) = B_t(-x,y),
\end{align}
where $B_t(\bm r) = B(\bm r) = \hat z \cdot \nabla \times \bm A(\bm r)$.

If we now further neglect the out-of-plane displacements, which is well justified for angles $\theta > 1^\circ$, then $\bm A_b(\bm r) = -\bm A_t(\bm r)$ at every point. Here we used that $\bm u_b(\bm r) = -\bm u_t(\bm r)$ for heterostrain. In this case, the first-star approximation yields
\begin{equation} \label{eq:pmf}
    \bm A(\bm r) = \frac{3B_0L^2}{8\pi^2} \sum_{i=1}^3 \hat z \times \bm g_i \cos \left( \bm g_i \cdot \bm r \right).
\end{equation}
Indeed, this is the form that we obtain if we plug in the first-star approximation for the relative in-plane displacement field given by Eq. (\textcolor{blue}{4}) of the main text with
\begin{equation} \label{eq:B0}
    B_0 = \frac{h}{e} \frac{\alpha_1 \beta}{L^2} = \frac{h}{e} \frac{c_1 \beta}{a^2},
\end{equation}
which is independent of the twist angle \cite{ceferino2023pseudomagnetic}. Using 
$c_1 = 6.5 \times 10^{-5}$ and $\beta = 2$ we find $B_0 \approx 9 \, \text{T}$. Note that we have not taken into account the out-of-plane contribution to the strain tensor. This is justified for twists $\theta > 1^\circ$. In this case, the in-plane contribution to the strain tensor is proportional to $cL/a < 0.001$, while the out-of-plane contribution scales as $(\Delta h/L)^2$ where $\Delta h \sim 0.01 \, \text{\AA}$. The corresponding pseudo magnetic field (PMF) becomes
\begin{equation}
    B(\bm r) = -2B_0 \sum_{i=1}^3 \sin \left( \bm g_i \cdot \bm r \right),
\end{equation}
which vanishes at AA points and has extrema at AB and BA points given by $\pm 3 \sqrt{3} B_0 \approx \pm 46 \, \text{T}$, respectively. We can further define an effective magnetic length
\begin{equation}
    \ell_0 = \sqrt{\frac{\hbar}{eB_0}} = \frac{a}{\sqrt{2\pi c_1 \beta}} \approx 8.6 \, \text{nm}.
\end{equation}
We note that the effect of the PMF on the electronic structure is significant only when $L / \ell_0$ is large. We show the PMF calculated from the \textsc{lammps} data, including out-of-plane contributions, in Fig.\ \ref{fig:pmf}.

\begin{figure}
    \centering
    \includegraphics[width=.95\linewidth]{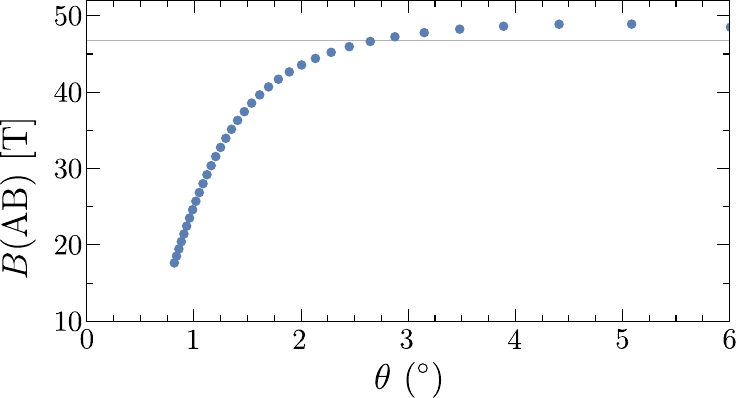}
    \caption{Pseudo magnetic field from heterostrain calculated with \textsc{lammps} at the Bernal stacking point $\bm r_\text{AB} = (L/\sqrt{3},0)$, indicated by the black dot in Fig.\ \ref{fig:pmf}, as a function of twist angle. The vertical line gives the estimate from Eq.\ \eqref{eq:B0}. Here we used $20$ reciprocal stars for the displacement fields.}
    \label{fig:pmfAB}
\end{figure}

The value for the strength of the PMF is one order of magnitude smaller than reported in Ref.\ \cite{nam2017lattice}. This difference is attributed to the fact that the value for $c_1 = V_1 / \mu$ in Ref.\ \cite{nam2017lattice} is about ten times larger.
Using molecular dynamics simulations, we estimate $c_1 \approx 6.5 \times 10^{-5}$ for tBG. Moreover, the PMF does not vary significantly with twist angle for $\theta > 2^\circ$. This follows from the scaling $(L / a)^2$ of the in-plane displacement field in lowest order. Since each spatial derivative brings down a factor $a / L$, the magnitude of the PMF becomes independent of twist angle. Deviations arise due to relatively small out-of-plane contributions or in-plane contributions that scale as $1/\sin^4(\theta/2)$ and start to become appreciable for smaller angles. We further find that the PMF decreases monotonically as the twist angle is reduced, as shown in Fig.\ \ref{fig:pmfAB}. This is similar to Ref.\ \cite{nam2017lattice} albeit at a smaller twist angle for the same reason as outlined above.

The decrease in the magnitude of the PMF can be understood as follows. First, from symmetry it follows that a PMF induced by in-plane heterostrain vanishes along domain walls separating AB and BA regions:
\begin{equation}
    B(x,y) = -B(-x,y),
\end{equation}
because $\mathcal C_{2y}$ exchanges the valleys, layers, and flips the direction of $\hat z$. Moreover, as the twist angle decreases, domains with nearly uniform AB and BA stacking start to form. While the strain is nonzero in these regions, the PMF is expected to decrease regardless because the microscopic $\mathcal C_{3z}$ symmetry is locally restored in these domains. Hence the PMF is pushed entirely to a narrow region surrounding the domain walls.

\section{Fermi velocity} \label{app:velocity}

\begin{figure}
    \centering
    \includegraphics[width=\linewidth]{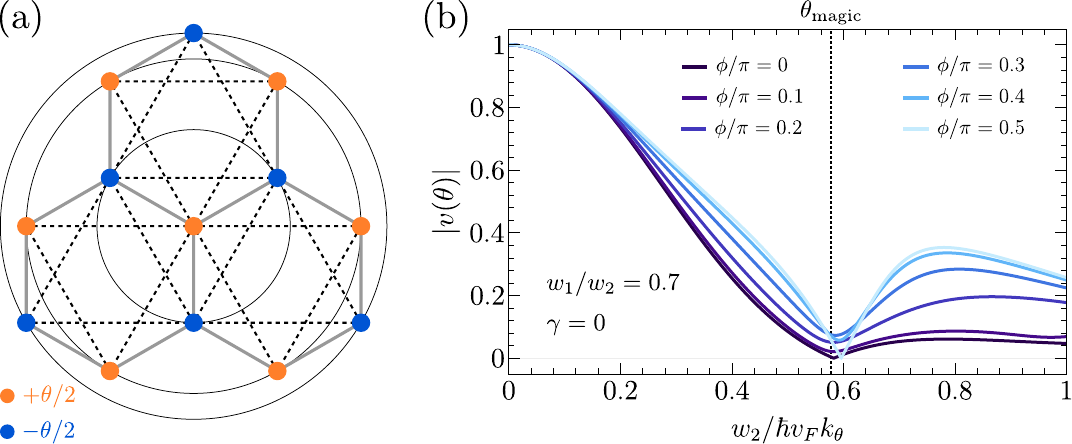}
    \caption{(a) Moir\'e replicas of Dirac points at valley $K_+$ of the top (orange) and bottom (blue) layer. Links between Dirac points of different layers (solid lines) represent interlayer couplings in the first moir\'e harmonic, while links between the same layer (dashed lines) are coupling induced by the pseudo magnetic field. (b) Converged Fermi velocity for $w_2/w_1 = 0.7$ and $\zeta = 0$ for different values of $\phi$ as indicated. The horizontal dashed line corresponds to $w_2 / \hbar v_F k_\theta = 1/\sqrt{3}$.}
    \label{fig:perturbation}
\end{figure}

In this section, we obtain an approximate expression for the Fermi velocity by restricting the interlayer moir\'e coupling in Fourier space. In particular, we retain one copy of the top layer Dirac cone and three moir\'e copies related by $\mathcal C_{3z}$ of the bottom layer \cite{bistritzer2011moire}. This corresponds to taking the smallest circle with radius $k_\theta = 4\pi / 3L$ centered around the principal Dirac point of the top layer, as illustrated in Fig.\ \ref{fig:perturbation}(a). We also define the dimensionless parameters
\begin{equation}
    \alpha_{1,2} = \frac{w_{1,2}}{\hbar v_F k_\theta}, \quad 
    \zeta = \frac{1}{\sqrt{3} k_\theta^2 \ell_0^2} = \frac{3\sqrt{3}}{16\pi^2} \frac{L^2}{\ell_0^2}.
\end{equation}
Note that it is important to keep track of the rotation of the layers. Here we always consider the case where the top (bottom) layer is rotated by $+\theta/2$ ($-\theta/2$). In this case, the relative displacement field $\bm u(\bm r)$ due to lattice relaxation with $\bm u_{t/b} = \pm \bm u$ should have positive curl at AA regions and negative curl at AB regions. This is because under $\bm u \rightarrow -\bm u$ the pseudo magnetic field in a given valley changes sign, leading to the wrong conclusions.

In this approximation, the Bloch Hamiltonian for valley $K_+$ in dimensionless units becomes
\begin{widetext}
\begin{align}
    h(\bm k) & = \begin{bmatrix} 
    \bm k \cdot \bm \sigma & T_1 & T_2 & T_3 \\
    T_1^\dag & \left( \bm k - \bm q_1 \right) \cdot \bm \sigma & -\zeta \bm A_1 \cdot \bm \sigma & -\zeta \bm A_3 \cdot \bm \sigma \\
    T_2^\dag & -\zeta \bm A_1 \cdot \bm \sigma & \left( \bm k - \bm q_2 \right) \cdot \bm \sigma & -\zeta \bm A_2 \cdot \bm \sigma \\
    T_3^\dag & -\zeta \bm A_3 \cdot \bm \sigma & -\zeta \bm A_2 \cdot \bm \sigma & \left( \bm k - \bm q_3 \right) \cdot \bm \sigma
    \end{bmatrix} \\
    & \equiv \begin{bmatrix} 
    0 & T_1 & T_2 & T_3 \\
    T_1^\dag & -h_1 & -A_{12} & -A_{13} \\
    T_2^\dag & -A_{21} & -h_2 & -A_{23} \\
    T_3^\dag & -A_{31} & -A_{32} & -h_3
    \end{bmatrix} + \begin{bmatrix} 
    \bm k \cdot \bm \sigma & 0 & 0 & 0 \\
    0 & \bm k \cdot \bm \sigma & 0 & 0 \\
    0 & 0 & \bm k \cdot \bm \sigma & 0 \\
    0 & 0 & 0 & \bm k \cdot \bm \sigma
    \end{bmatrix},
\end{align}
\end{widetext}
with $\bm q_1 = (0,-1)$, $\bm q_{2,3} = (\pm\sqrt{3}/2,1/2)$, and $\bm A_i = \hat z \times \hat g_i$ ($i=1,2,3$). Here we moved the momentum origin to the principal Dirac point of the top layer and defined $h_i = \bm q_i \cdot \bm \sigma$ and
\begin{equation}
    A_{ij} = \zeta \begin{pmatrix} 0 & \bm A_1 \cdot \bm \sigma & \bm A_3 \cdot \bm \sigma \\
    \bm A_1 \cdot \bm \sigma & 0 & \bm A_2 \cdot \bm \sigma \\
    \bm A_3 \cdot \bm \sigma & \bm A_2 \cdot \bm \sigma & 0
    \end{pmatrix}.
\end{equation}

Following Ref.\ \cite{bistritzer2011moire}, the Fermi velocity is obtained from perturbation theory in $|\bm k|$. Hence, we first need to diagonalize $h(\bm k)$ at the origin. To this end, we first write the wave function at $|\bm k| = 0$ as $\Psi = \left( \psi_0, \psi_1, \psi_2, \psi_3 \right)^t$. Here $\psi_0$ corresponds to the top layer and $\psi_i$ ($i=1,2,3$) are the components of the bottom layer at momentum $\bm q_i$. The eigenvalue equation can then be written as
\begin{align}
    \sum_{i=1}^3 T_i \psi_i & = \varepsilon \psi_0, \label{eq:approx1} \\
    T_i^\dag \psi_0 - \sum_{j=1}^3 \left( h_i \delta_{ij} + A_{ij} \right) \psi_j & = \varepsilon \psi_i. \label{eq:approx2}
\end{align}
Solving Eq.\ \eqref{eq:approx2} for the components of the bottom layer and plugging this into Eq.\ \eqref{eq:approx1} yields
\begin{widetext}
\begin{equation} \label{eq:eval}
    \begin{pmatrix} T_1 & T_2 & T_3 \end{pmatrix} \begin{pmatrix} \varepsilon + h_1 & A_{12} & A_{13} \\ A_{21} & \varepsilon + h_2 & A_{23} \\ A_{31} & A_{32} & \varepsilon + h_3 \end{pmatrix}^{-1} 
    \begin{pmatrix} T_1^\dag \\ T_2^\dag \\ T_3^\dag
    \end{pmatrix} \psi_0 = \varepsilon \psi_0.
\end{equation}
There are two types of solutions. Either $\psi_0$ is nonzero with doubly-degenerate eigenvalues and solutions $\psi_0^{(1)} = (1,0)^t$ and $\psi_0^{(2)} = (0,1)^t$, or otherwise $\psi_0$ vanishes. In the former, the components from the second layer are obtained from Eq.\ \eqref{eq:approx2}. We then find that the matrix on the left-hand side of Eq.\ \eqref{eq:eval} is proportional to the unit matrix, and the energies are given by the roots of a depressed cubic:
\begin{equation} \label{eq:pol}
    \varepsilon \left[ \left( \varepsilon + \zeta + 1 \right) \left( \varepsilon - \zeta - 1 \right) - 3 \left( w_1^2 + w_2^2 \right) \right] - 6 w_1 w_2 \left( 1 + \zeta \right) \sin \phi = 0,
\end{equation}
where $\phi$ is the phase of the AA coupling. 
However, this equation only yields six eigenvalues. The remaining two eigenvalues are solutions with $\psi_0 = (0,0)^t$. We henceforth focus on solutions $\Psi_{1,2}$ with nonzero $\psi_0$ and the smallest $|\varepsilon|$. 
The Fermi velocity is obtained by projecting $h(\bm k)$ on these eigenstates. In the basis $\{ \psi_0^{(1)}, \psi_0^{(2)} \}$ we have
\begin{equation}
    \begin{pmatrix} 0 & v \\ v^*& 0 \end{pmatrix} = \frac{\sigma_x + \begin{pmatrix} T_1 & T_2 & T_3 \end{pmatrix} \begin{pmatrix} \varepsilon + h_1 & A_{12} & A_{13} \\ A_{21} & \varepsilon + h_2 & A_{23} \\ A_{31} & A_{32} & \varepsilon + h_3 \end{pmatrix}^{-1} \begin{pmatrix} \sigma_x & 0 & 0 \\ 0 & \sigma_x & 0 \\ 0 & 0&  \sigma_x \end{pmatrix} \begin{pmatrix} \varepsilon + h_1 & A_{12} & A_{13} \\ A_{21} & \varepsilon + h_2 & A_{23} \\ A_{31} & A_{32} & \varepsilon + h_3 \end{pmatrix}^{-1}
    \begin{pmatrix} T_1^\dag \\ T_2^\dag \\ T_3^\dag
    \end{pmatrix}}{1 + \psi_0^{{(1)}\dag} \begin{pmatrix} T_1 & T_2 & T_3 \end{pmatrix} \begin{pmatrix} \varepsilon + h_1 & A_{12} & A_{13} \\ A_{21} & \varepsilon + h_2 & A_{23} \\ A_{31} & A_{32} & \varepsilon + h_3 \end{pmatrix}^{-2} 
    \begin{pmatrix} T_1^\dag \\ T_2^\dag \\ T_3^\dag
    \end{pmatrix} \psi_0^{(1)}},
\end{equation}
\end{widetext}
with $v$ in units of $v_F$. In the special case $\phi = 0$, there is a zero-energy solution $\varepsilon=0$ and we find 
\begin{equation}
    v(\theta) = \frac{\left( 1 + \zeta \right)^2 - 3 \alpha_2^2}{\left( 1 + \zeta \right)^2 + 3 \alpha_1^2 + 3 \alpha_2^2},
\end{equation}
which vanishes for $\alpha_2 = \left( 1 + \zeta \right) / \sqrt{3}$. Hence, using that $k_\theta \approx \theta 4\pi / 3a$, we obtain
\begin{equation}
    \frac{16 \pi^2}{3\sqrt{3}} \, \theta^2 - \frac{4 \pi w_2 a}{\hbar v_F} \, \theta + \left( \frac{a}{\ell_0} \right)^2 = 0,
\end{equation}
which has one physical solution given by
\begin{align}
    \theta_\text{magic} & = \frac{3\sqrt{3}w_2a}{8\pi\hbar v_F} + \frac{\sqrt{3}}{8\pi} \sqrt{\left( \frac{3aw_2}{\hbar v_F} \right)^2 - \frac{4\sqrt{3}}{\ell_0^2}} \\
    & = \frac{3 \sqrt{3} w_2 a}{4\pi \hbar v_F} - \frac{\hbar v_F a}{4 \pi w_2 \ell_0^2} + \mathcal O( a^4 / \ell_0^4 ).    
\end{align}
Thus the pseudo magnetic field merely shifts the magic angle to a lower value. However, when $\phi$ is nonzero, the velocity becomes complex. Now it becomes impossible to tune both the real and imaginary part to zero with a single parameter. We confirm this numerically, as shown in Fig.\ \ref{fig:perturbation}(b). For small $\phi$, the magnitude of the velocity at $\theta_\text{magic}$ is approximately given by
\begin{equation}
    |v_\text{min}| = \frac{4 w_1^2 w_2^2 \phi}{\left( w_1^2 + 2 w_2^2 \right)^2}.
\end{equation}

\section{Moir\'e potential} \label{app:mpotential}

In this section, we estimate the effect of lattice relaxation on the (local) moir\'e potential. For the rigid case, we have $w_1 = w_2 = w_0$ where $w_1$ and $w_2$ are the tunneling amplitudes of the first moir\'e star between equal and opposite sublattices, respectively. In this case, the corresponding moir\'e potentials are related by a translation. However, lattice relaxation breaks this symmetry leading to a difference in the moir\'e potentials which varies as a function of twist angle. Below, we use the general definition of the moir\'e potential \cite{kang2023,kang2023b} to obtain an approximate expression for $w_1$ and $w_2$ as a function of twist angle using our theory for lattice relaxation valid for twist angles $\theta \gtrsim 1^\circ$.

The local moir\'e coupling is given by
\begin{equation}
    H_\text{inter} = \sum_{\sigma\sigma'} \int d^2\bm x \, \psi_{\sigma1}^\dag T_{\sigma\sigma'}(\bm x) \psi_{2\sigma'}(\bm x) + \text{h.c.},
\end{equation}
with $T_{\sigma\sigma'}(\bm x)$ the moir\'e potential between sublattice $\sigma$ on layer 1 and sublattice $\sigma'$ on layer 2. If we only consider acoustic displacements in the absence of homostrain, then in the two-center approximation \cite{kang2023,kang2023b},
\begin{align}
    T_{\sigma\sigma'}(\bm x) & = \sum_{\bm b} e^{i \bm b \cdot \bm \delta_{\sigma\sigma'}} e^{i \left( \bm b + \bm K \right) \cdot \bm \phi(\bm x)} \breve t[\bm b + \bm K, h(\bm x)],
\end{align}
where the sum runs over reciprocal lattice vectors $\bm b$ of the monolayer and $\bm \delta_{\sigma\sigma'} = \bm \delta_\sigma - \bm \delta_{\sigma'}$ with $\bm \delta_\sigma$ the sublattice position in the graphene unit cell. We further introduced
\begin{equation}
    \bm \phi(\bm x) = \frac{a}{L} \hat z \times \bm x + 2\bm u(\bm x),
\end{equation}
the local stacking configuration and the Fourier transform of the interlayer hopping amplitude
\begin{equation}
    \breve t(\bm k, h) = \frac{1}{A_\text{g}} \int d^2\bm y \, e^{-i \bm k \cdot \bm y}  t_\perp \left[ \bm y + 2h \hat z \right],
\end{equation}
with $A_\text{g} = \sqrt{3} a^2/2$ the unit cell area of monolayer graphene. These Fourier components are real and only depend on $|\bm k|$ in the two-center approximation since $\breve t(R\bm k, h) = \breve t(\bm k, h)$. Further note that
\begin{equation}
    e^{i \left( \bm b + \bm K \right) \cdot \bm \phi(\bm x)} = e^{i \left( \bm g + \bm q_1 \right) \cdot \bm x} e^{i \left( \bm b + \bm K \right) \cdot \bm u(\bm x)}
\end{equation}
with $\bm q_1 = \tfrac{a}{L} \bm K \times \hat z$ and $\bm g = \tfrac{a}{L} \bm b \times \hat z$. If we now approximate $2h(\bm x)$ with the average interlayer distance $2h_0$, we obtain
\begin{equation}
    T_{\sigma\sigma'}(\bm x) = \sum_{\bm q} e^{i \bm q \cdot \bm x} \breve T_{\sigma\sigma'}(\bm q),
\end{equation}
with $\bm q = \bm g + \bm q_1$ and
\begin{widetext}
\begin{align} \label{eq:mpot}
    \breve T_{\sigma\sigma'}(\bm q) & =
    \sum_{\bm b'} e^{i \bm b' \cdot \bm \delta_{\sigma\sigma'}} \breve t_\perp(\bm b' + \bm K, h_0) \left[ \frac{1}{A_\text{m}} \int_{\text{moir\'e cell}} d^2 \bm x \, e^{-i \bm q \cdot \bm x} e^{i ( \bm b' + \bm K ) \cdot \bm \phi(\bm x)} \right] \\
    & = \sum_{\bm b'} e^{i \bm b' \cdot \bm \delta_{\sigma\sigma'}} \breve t_\perp(\bm b' + \bm K, h_0) \left[ \frac{1}{A_\text{m}} \int_{\text{moir\'e cell}} d^2 \bm x \, e^{-i ( \bm g - \bm g' ) \cdot \bm x} e^{2i ( \bm b' + \bm K ) \cdot \bm u(\bm x)} \right].
\end{align}
\end{widetext}

\subsection{Symmetry constraints}

Using the above form, we find relations between Fourier components of symmetry-related $\bm q$ vectors. We start with $120^\circ$ degree rotations about the $z$ axis. After a change of variable in the integral, we obtain (dropping the prime in the sum for notational convenience)
\begin{widetext}
\begin{align}
    \breve T_{\sigma\sigma'}(\mathcal C_{3z} \bm q) & = \frac{1}{A_\text{m}} \int_{\text{moir\'e cell}} d^2 \bm x \, e^{-i \bm q \cdot \bm x} \sum_{\bm b} e^{i \bm b \cdot \bm \delta_{\sigma\sigma'}} e^{i \mathcal C_{3z}^{-1} \left( \bm b + \bm K \right) \cdot \bm \phi(\bm x)} \breve t_\perp(\bm b + \bm K, h_0) \\
    & = \frac{e^{i \left( \mathcal C_{3z} \bm K - \bm K \right) \cdot \bm \delta_{\sigma\sigma'}}}{A_\text{m}} \int_{\text{moir\'e cell}} d^2 \bm x \, e^{-i \bm q \cdot \bm x} \sum_{\bm b} e^{i \bm b \cdot \bm \delta_{\sigma\sigma'}} e^{i \left( \mathcal C_{3z} \bm b - \bm b \right) \cdot \bm \delta_{\sigma\sigma'}} e^{i \left( \bm b + \bm K \right) \cdot \bm \phi(\bm x)} \breve t_\perp \left[ \mathcal C_{3z} \left( \bm b + \bm K \right), h_0 \right],
\end{align}
\end{widetext}
where we let $\bm b \rightarrow \mathcal C_{3z} \bm b + \mathcal C_{3z} \bm K - \bm K$ since the sum runs over all reciprocal vectors. Moreover, since $e^{i \left( \mathcal C_{3z} \bm b - \bm b \right) \cdot \bm \delta_{\sigma\sigma'}} = e^{i \bm b \cdot ( \mathcal C_{3z}^{-1} \bm \delta_{\sigma\sigma'} - \bm \delta_{\sigma\sigma'} )} = 1$, we obtain
\begin{equation}
    \breve T_{\sigma\sigma'}(\mathcal C_{3z} \bm q) = e^{i \left( \mathcal C_{3z} \bm K - \bm K \right) \cdot \bm \delta_{\sigma\sigma'}} \breve T_{\sigma\sigma'}(\bm q).
\end{equation}
Secondly, we consider $\mathcal C_{2y}$ symmetry where we assume that the original monolayers are oriented with the zigzag direction along the $x$ axis. We find
\begin{widetext}
\begin{align}
    \breve T_{\sigma\sigma'}(\mathcal C_{2y} \bm q) & = \frac{1}{A_\text{m}} \int_{\text{moir\'e cell}} d^2 \bm x \, e^{-i \bm q \cdot \bm x} \sum_{\bm b} e^{i \bm b \cdot \bm \delta_{\sigma\sigma'}} e^{-i \mathcal C_{2y} \left( \bm b + \bm K \right) \cdot \bm \phi(\bm x)} \breve t_\perp(\bm b + \bm K, h_0) \\
    & = \frac{e^{-i \left( \mathcal C_{2y} \bm K + \bm K \right) \cdot \bm \delta_{\sigma\sigma'}}}{A_\text{m}} \int_{\text{moir\'e cell}} d^2 \bm x \, e^{-i \bm q \cdot \bm x} \sum_{\bm b} e^{-i \bm b \cdot \mathcal C_{2y} \bm \delta_{\sigma\sigma'}} e^{i \left( \bm b + \bm K \right) \cdot \bm \phi(\bm x)} \breve t_\perp \left[ -\mathcal C_{2y} \left( \bm b + \bm K \right), h_0 \right]
\end{align}
\end{widetext}
where we changed the integration variable, used that $\bm \phi(\mathcal C_{2y} \bm x) = -\mathcal C_{2y} \bm \phi(\bm x)$, and let $\bm b \rightarrow -\mathcal C_{2y} ( \bm b + \bm K ) - \bm K$ in the last line. Also note that $\left( \mathcal C_{2y} \bm K + \bm K \right) \cdot \bm \delta_{\sigma\sigma'} = 0$ and $\mathcal C_{2y} \bm \delta_{\sigma\sigma'} = \bm \delta_{\sigma\sigma'} + \bm a$. Hence, we find
\begin{equation}
    \breve T_{\sigma\sigma'}(-q_x, q_y) = \breve T_{\sigma'\sigma}(\bm q) = \left[ \breve T_{\sigma\sigma'}(\bm q) \right]^*, 
\end{equation}
where the last equality follows from $\mathcal C_{2z}$ and the reality of $\breve t_\perp(\bm b + \bm K, h_0)$. We find numerically that all these conditions are met. Therefore the reality of $\breve t_\perp$, which follows from the two-center approximation for the interlayer tunneling amplitude, implies that $\breve T_\text{AA}(\bm q)$ is real.

\subsection{Lowest-order corrections}

The Fourier components of the interlayer hopping amplitude are dominated by $\bm b + \bm K = \{ \bm K, \mathcal C_{3z} \bm K, \mathcal C_{3z}^2 \bm K \}$. If we only consider these contributions, we obtain
\begin{widetext}
\begin{equation}
    T_{\sigma\sigma'}(\bm x) \simeq w_0 \left[ e^{i \bm K \cdot \bm \phi(\bm x)} + e^{i ( \mathcal C_{3z} \bm K - \bm K ) \cdot \bm \delta_{\sigma\sigma'}} e^{i \mathcal C_{3z} \bm K \cdot \bm \phi(\bm x)} +e^{i ( \mathcal C_{3z}^2 \bm K - \bm K ) \cdot \bm \delta_{\sigma\sigma'}} e^{i \mathcal C_{3z}^2 \bm K \cdot \bm \phi(\bm x)} \right],
\end{equation}
\end{widetext}
where $w_0 = \tilde t_\perp(\bm K,h_0)$ which depends weakly on the twist angle through $h_0$. Since $e^{i \bm K \cdot \bm \phi(\bm x)} = e^{i \bm q_1 \cdot \bm r} e^{2i\bm K \cdot \bm u(\bm x)}$, we recover the familiar moir\'e potential in the absence of relaxation. In general in Eq.\ \eqref{eq:mpot} we expand
\begin{equation}
    e^{i(\bm b + \bm K) \cdot \bm u(\bm x)} \simeq 1 + 2i (\bm b + \bm K) \cdot \bm u(\bm x),
\end{equation}
which yields
\begin{widetext}
\begin{equation}
    \breve T_{\sigma\sigma'}(\bm q) \simeq e^{i \bm b \cdot \bm \delta_{\sigma\sigma'}} \breve t_\perp(\bm b + \bm K, h_0) + \sum_{\bm b'} e^{i \bm b' \cdot \bm \delta_{\sigma\sigma'}} \breve t_\perp(\bm b' + \bm K, h_0) \left( \bm b' + \bm K \right) \cdot 2i\bm u_{\bm g - \bm g'},
\end{equation}
\end{widetext}
with $\bm b = \tfrac{L}{a} \hat z \times (\bm q - \bm q_1)$. Let us consider the first moir\'e star with $\bm q = \bm q_1$ and thus $\bm b = \bm g = \bm 0$. We then have approximately
\begin{equation}
    \breve T_{\sigma\sigma'}(\bm q_1) \simeq w_0 \left[ 1 + \sum_{n=1}^2 e^{i \bm b_n \cdot \bm \delta_{\sigma\sigma'}} \left( \bm b_n + \bm K \right) \cdot 2i\bm u_{- \bm g_n} \right],
\end{equation}
where $|\bm b_n + \bm K| = |\bm K|$ with $\bm b_n$ nonzero since $\bm u_{\bm 0} = \bm 0$. Using our results for the displacement field from Eq.\ 4 of the main text, we find
\begin{widetext}
\begin{align}
    \breve T_{\sigma\sigma'}(\bm q_1) & \simeq  w_0 \left[ 1 - \frac{a \alpha_1}{L} \frac{3L^2}{8\pi^2} \sum_{i=1}^2 e^{i \bm b_n \cdot \bm \delta_{\sigma\sigma'}} \left( \bm b_n + \bm K \right) \cdot \left( \hat z \times \bm g_n \right) \right] \\
    & = w_0 \left[ 1 - 2\alpha_1 \sum_{n=1}^2 e^{i \bm b_n \cdot \bm \delta_{\sigma\sigma'}} \left( 1 + \frac{\hat K \cdot \hat b_n}{\sqrt{3}} \right) \right].
\end{align}
\end{widetext}
Thus, using Eq.\ \eqref{eq:Salpha1}, we find that in lowest order the first moir\'e shell is modified as
\begin{align}
    \breve T_\text{AA}(\bm q_1) & \simeq  w_0 \left( 1 - \frac{c_1}{2 \sin^2 \tfrac{\theta}{2}} \right), \label{eq:w1AA} \\
    \breve T_\text{AB}(\bm q_1) & \simeq  w_0 \left( 1 + \frac{c_1}{4 \sin^2 \tfrac{\theta}{2}} \right), \label{eq:w1AB}
\end{align}
and the third moir\'e shell with $|\bm q| = \sqrt{7}k_\theta$ is generated by lattice relaxation:
\begin{equation} \label{eq:w3}
    \breve T_{\sigma\sigma'}(3\bm q_1 + \bm q_2) \simeq \frac{w_0 c_1}{4 \sin^2 \tfrac{\theta}{2}},
\end{equation}
with $\bm q_2 = \mathcal C_{3z} \bm q_1$ and which is independent of the sublattice indices.
\begin{figure} 
    \centering\includegraphics[width=\linewidth]{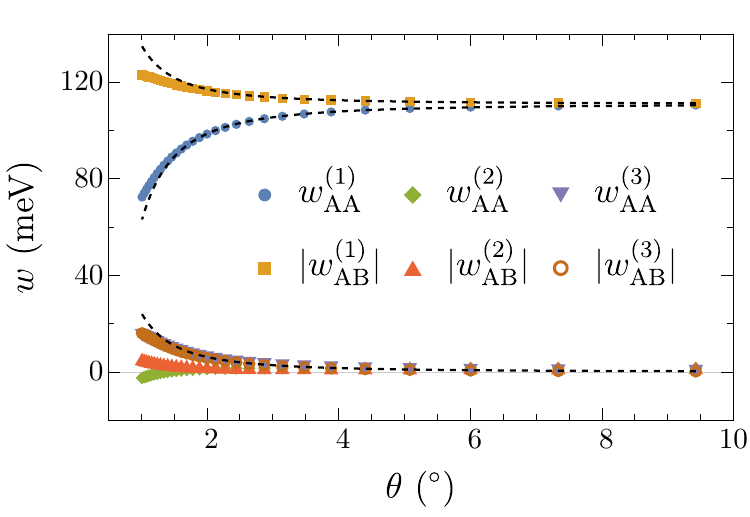}
    \caption{Comparison of calculated interlayer coupling parameters using \textsc{lammps} together with a Slater-Koster form for the hopping amplitudes (data points) with the theory (dashed lines) from Eqs.\ \eqref{eq:w1AA}, \eqref{eq:w1AB}, and \eqref{eq:w3}.}
    \label{fig:w}
\end{figure}

We compare these results to numerical calculations using \textsc{lammps} data together with a Slater-Koster parameterization \cite{moon2013_optical} for the hopping amplitudes with fixed $h_0 = 3.35 \; \text{\r A}$. This is shown in Fig.\ \ref{fig:w}. Here we set $c_1 = 6.8 \times 10^{-5}$ and $w_0 = 110.89 \, \text{meV}$.
The numerical results were obtained by numerically computing the inverse Fourier transform of Eq.\ \eqref{eq:mpot} where $w_1$ and $w_2$ are the components of the first star for the AA and AB tunneling amplitudes, respectively.

\subsection{First-star approximation with relaxation parameter $\phi$}
\label{sectionOldFig}

As we show in \ref{phisymmetry}, the relaxation parameter $\phi$ is allowed by symmetry. Previous numerical results from Ref. \cite{carr2019exact} showed that $\phi$ has a small non-zero value and that higher-order shells beyond the first star become more pronounced with relaxation. Therefore, both effects contribute to the reconstruction of the electronic spectrum of tBG. However, in this work, we introduce an alternative approach (see Equation \textcolor{blue}{8} in the main text) that employs only the first star while using $\phi$ to encode information about all other relaxation effects, which, in principle, could be captured only when higher-order shells are considered. This proposed model can, to an excellent degree, match the spectrum of the relaxed tight-binding tBG electronic model.

\begin{figure*}[!t]
    \centering
    \includegraphics[width=1.\textwidth]{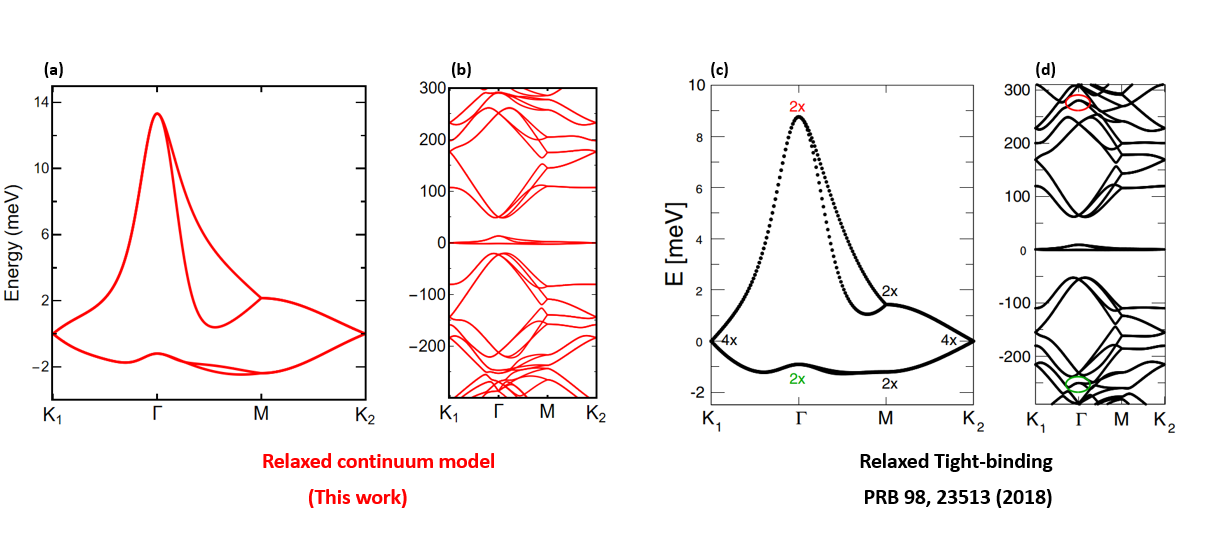}
    \caption{Validating the effective first-star approximation with the $\phi$ parameter against relaxed tight-binding calculations. (a) Low-energy bands of the proposed continuum model for $\theta = 1.08^\circ$. (b) Zoom-out view of the band structure of the effective model showing additional bands around charge neutrality points. (c, d) Corresponding tight-binding results to (a) and (b) taken from \cite{angeli2018emergent}. It is evident that the effective continuum model captures not only the low-energy bands but also the high-energy features of the spectrum. Similar agreement is achieved in comparison with the results of \cite{leconte2022relaxation} (not shown here).}
    \label{figphi}
\end{figure*}

Figure \ref{figphi} demonstrates the validation of both the low and high-energy features of the proposed effective relaxed electronic continuum model against tight-binding calculations that include atomic relaxation \cite{angeli2018emergent,leconte2022relaxation}. The relaxation parameter $\phi$ aids in reproducing the relaxed tight-binding energy spectrum, and the electron-hole asymmetry is accurately captured. The excellent agreement with the tight-binding calculations indicates that our proposed effective relaxed model can serve as a robust non-interacting Hamiltonian, upon which interacting theories could be developed.

\section{Molecular Dynamics Simulations}

In the context of small twist angles where the moir\'e structure is notably large and contains a substantial number of atoms within the supercell, conducting first-principle calculations proves to be prohibitively expensive. In this regime, we calculate atomic relaxation with molecular dynamics simulations using the Large-scale Atomic/Molecular Massively Parallel Simulator (\textsc{lammps}) code which employs classical interatomic force field models \cite{THOMPSON2022108171}. While these molecular dynamics simulations allow for larger supercell sizes, they have inherent limitations on accuracy over the choice of interatomic potentials. It is our experience that while different interatomic potentials might give slightly different numerical values for the scaling factor $c_1$, their qualitative behavior and symmetry properties are identical. For twisted bilayer graphene, we use the Drip potential for interlayer interactions and the REBO potential for intralayer interactions \cite{Donald:WBrenner_2002,leconte2022relaxation,JeilFF}. For twisted WSe$_2$ we use the KC potential for interlayer interactions and the SW potential for intralayer interactions with SW/mod style \cite{doi:10.1021/acs.jpcc.8b10392, Jiang_2015}. For this work, we perform relaxation calculations for commensurate twist angles ranging from $\theta = 0.05^\circ$ to $\theta = 20^\circ$. The smallest twist angles correspond to moir\'e cells with over $4.3$ million atoms.  Despite the large number of atoms in the simulation cell, the geometry optimizations remain computationally tractable due to the low cost of the classical potentials.

\end{document}